\documentclass[12pt]{elsarticle}

\usepackage{graphicx}
\usepackage{epsfig}
\usepackage{subcaption}
\usepackage{amssymb,amsmath}
\usepackage{float}
\usepackage{algorithm2e}
\usepackage{soul,xcolor}
\usepackage{multirow}
\usepackage{array}
\usepackage{todo}
\usepackage{hyperref}
\usepackage{nicefrac} 
\usepackage{booktabs} 
\usepackage[percent]{overpic}
\usepackage[left=1in,right=1in]{geometry}

\renewcommand{\vec}[1]{\mathbf{#1}}

\usepackage{pgfplots} 
\usepackage{tikz}

\begin{document}
\begin{frontmatter}

\title{Influence of capillary viscous flow on melting dynamics}

\author[iisc]{Michael Blank\corref{cor2}}

\author[iisc]{Prapanch Nair\corref{cor1}}
\ead{prapanch.nair@fau.de}

\author[iisc]{Thorsten P\"oschel\corref{cor2}}

\cortext[cor1]{Principal corresponding author}

\address[iisc]{ Institute for Multiscale Simulation, Friedrich-Alexander Universit\"at Erlangen-N\"urnberg, Erlangen.}

\begin{abstract}
  
  The rate of melting of a solid and the rate of deformation of the resulting 
  melt due to capillary forces are comparable in additive manufacturing 
  applications. 
  This dynamic structural change of a melting solid  
  is extremely challenging to study experimentally. Using meshless numerical simulations 
  we show the influence of the flow of the melt on the heat transfer and 
  resulting phase change.   
  
  We introduce an accurate and robust Incompressible Smoothed Particle 
  Hydrodynamics method to simulate melting of solids and the ensuing 
  fluid-solid interaction. 
  We present validation for the heat transfer
  across free surface and the melting interface evolution, separately.
  We then present two applications for this coupled multiphysics simulation method---
  the study of rounding of an arbitrarily 
  shaped particle during melting and the non-linear structural evolution of 
  three spheres undergoing agglomeration. In both the studies we use realistic
  transport and thermal properties for the materials so as to demonstrate 
  readiness of the method for  solving engineering problems in additive manufacturing.  
\end{abstract}

\begin{keyword}
Additive manufacturing \sep Incompressible Smoothed Particle Hydrodynamics \sep phase change \sep latent heat 
\sep melting dynamics

\end{keyword}

\end{frontmatter}

\section{Introduction}
\label{intro}

The evolution of topology in melting system, such as encountered in 
additive manufacturing applications, is often studied by decoupling the 
time scales of the flow of the melt and the phase change. Either static material 
properties are assumed, approximating the solids as highly viscous
fluids \cite{kirchhof2009three} or assuming no flow following melting \cite{stewartson1976stefan}. As the scope of additive manufacturing widens, 
there is an increasing 
need to resolve the tight coupling between flow of the melt 
and the phase change in order to achieve expected strength of the material.
For serious engineering applications, uniformity in 
density and strength of the manufactured parts demand a micro mechanical 
understanding of the structural evolution during successive melting 
and solidification the material undergoes several times. 
Experimental studies at 
such spatial and temporal resolutions may be expensive \cite{guo2013additive}, not to mention 
the high temperature conditions in which measurements need to be made. 
Traditional numerical approaches 
based on mesh based methods (Finite Volume Method, Finite Element Method) are 
often challenged by complex geometry evolutions;
this is inherently due to the need to maintain the neighbor connectivity information
between the computational nodes. Meshless methods hold several advantages for such multi-physics applications. 

Smoothed Particle Hydrodynamics (SPH) is a meshfree updated-Lagrangian method which 
was introduced by Ginghold and Monaghan \cite{gingold1977} and Lucy \cite{Lucy1977}
for treating astrophysical phenomena and gas dynamics. Since the method is
free from the need for the mesh connectivity information, many problems with 
complex interfaces and discontinuities in the domain are effectively solved 
by the method, making it a practical tool for multiphysics problems. 
For example, Monaghan presented an impressive SPH simulation of a methane gas bubble
toppling a trawler to explain the sinking of ships in Witches Hole in the 
North sea \cite{may2003can}. 
This multiphysics simulation coupled two phase flow with free surface to an 
interacting rigid body. Heat transfer \cite{morris1997modeling}, phase change \cite{farrokhpanah2017new},
and capillary flows \cite{nair2018dynamic} are solved separately using the method
and are being continously applied to a wide variety of engineering problems.

To successfully couple heat transfer, phase change and capillary flows in the
context of additive manufacturing applications, each of these modules 
need to be improved and validated for appropriate boundary conditions (BC).  Heat
transfer is solved in SPH using the passive scalar approach for temperature \cite{morris1997modeling}. 
The definition of accurate second order derivatives with discontinous transport 
coefficients makes it applicable to realistic problems. Heat transfer from 
an ambient constant (or time varying) temperature needs to be 
implemented as a Dirichlet BC for temperature. A thin layer of particles 
may be used for such BCs, but this would deteriorate the order
of accuracy of the method itself. Hence, for problems undergoing continous 
deformation, we present a free surface Dirichlet BC that is quite similar to the 
free surface Dirichlet BC for pressure in an earlier work \cite{nair2014improved}. 

The transport of passive scalar---through convection due to capillary flow
and conduction---causes phase
change based on the latent heat capacity of the material. Using SPH,
static phase change has been solved \cite{farrokhpanah2017new} using fixed particles for boundaries. Enthalpy based formulations have also been used for solving
heat transfer in SPH \cite{farrokhpanah2017new} as an alternative approach. However these involve 
computations of higher order derivatives than the temperature based approach. 
Capillary forces can be implemented in SPH using either continuum surface force
based models, geometric reconstruction of the interfaces or using pairwise 
potential forces. For free surfaces we find the potential forces to be more 
amenable \cite{nair2018dynamic} for free surfaces. Other capillary force
models may be implemented in this context. 

In Sec. \ref{model} of this paper we present the heat transfer, phase change and 
capillary flow models. In Sec. \ref{implementation} we present implementation of 
these modelts to the ISPH method. In this section we introduce our boundary 
condition model for heat transfer across the free surface. We also separately 
validate the heat tranfer across free surface and melting across the free surface
for different latent heat values in 2D and 3D. In Sec. \ref{results} we present applications
of this method to representative problems in additive manufacturing. We 
present the simulations of melting 
of complex shaped particles, compare the melting rate with available  
theoretical results for spherical shape to appreciate the effect of shape 
the ensuing viscous flow is coupled to the heat transfer problem. 
Finally, we simulate the agglomeration of a 
chain of three solid spheres undergoing melting with two different sets of 
material properties to show the the shape evolution is a result of instabilities
relating to the solid liquid interaction during the melting process.

\section{Governing Equations}
\label{model}
In the present study we assume the fluid (melt)
and the solid to be incompressible, with negligible density variation duringthe  
phase change. The transient heat transfer in the system is governed by the
enthalpy equation,
\begin{equation}
  \frac{\partial H}{\partial t} = \nabla \cdot (k\nabla T) ,
  \label{eq:enthalpy}
\end{equation}
where $H$ is the enthalpy, $k$ is the conductivity of the material and $T$ the temperature.
For an incompressible medium, the enthalpy $H$ can be written as a function of temperature. Therefore,
\begin{equation}
  \frac{dH}{dT}\frac{\partial T}{\partial t} = \nabla \cdot (k\nabla T).
  \label{eq:heat}
\end{equation}
Here $dH/dT$ is the specific heat capacity at constant pressure, $C$.
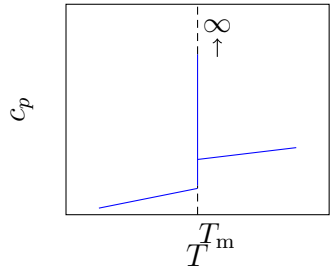
\begin{figure}[!htb]
\begin{center}
	\begin{tikzpicture}
	\begin{axis}[ 
	width=2in,
	xmin = 0, xmax =4,
	ymin=0, ymax=8,
	xticklabels={,,},
	yticklabels={,,},
	ticks=none,
	x label style={at={(axis description cs:0.5,0.1)},anchor=north},
	y label style={at={(axis description cs:0.1,0.5)},anchor=north},
	xlabel = $T$,
	ylabel = $c_{p} $,
	legend pos = north east,
	legend style={draw=none}
	]
	\addplot[samples=2, draw=blue, domain=0.5:2] {0.5*x+0};
	\addplot +[mark=none, draw=blue] coordinates {(2, 1.1) (2, 6.1)};
	\addplot[mark size = 1.0pt, draw=blue, domain=2:3.5] {0.3*x+1.5};
	\addplot +[mark=none, draw=black, dash pattern={on 3pt off 2pt}] coordinates {(2, 6.1) (2, 8)};
	\addplot +[mark=none, draw=black,dash pattern={on 3pt off 2pt}] coordinates {(2, 0) (2, 1.1)};
	\draw[black,->](axis cs:2.3,6.0)--(axis cs:2.3,6.7);
	\pgfplotsset{
		after end axis/.code={
			\node[black] at (axis cs:2.3,7.2){\small{$\infty$}};    			
			\node[black] at (axis cs:2.3,-0.8){\small{$T_{\mathrm{m}}$}};    		
		}
	}
	\end{axis}
	\end{tikzpicture}
	\caption{Increase of the specific heat capacity during the melting process.}
  \label{fig:spheatmodel}
\end{center}
\end{figure}

Phase change (melting and solidification) can be modelled using the above equation by defining an effective 
heat capacity \cite{farrokhpanah2017new} which increases (or decreases) 
by the 
latent heat of melting (or solidification), $L$, of the 
material at a melting (or solidifying) interface. This effective 
specific heat capacity can be defined as
\begin{equation}
  C_\textrm{eff} = \begin{cases}
    C_s & T<T_m \\
    C_m + L\delta (T-T_m) & T=T_m\\
    C_l & T>T_m
  \end{cases}
  \label{eq:effectiveheatcapacity}
\end{equation}
where $\delta (T-T_m) $ is the Dirac delta function, $C_s$ and $C_l$ are the
specific heat capacities of the material in the solid and liquid phases, respectively, 
and $C_m$ is the specific heat capacity of the material at the melting interface, at the melting temperature. 
This model visually represented in Fig. \ref{fig:spheatmodel}.
The specific heat capacity, in general, varies with temperature.

The liquid domain resulting from the melt is modelled here using a one-fluid formulation, neglecting the presence of 
air and is governed by the incompressible Navier Stokes equation given by:

\begin{equation}
  \frac{d\vec{u}}{dt} = -\frac{1}{\rho}\nabla P + \nabla \cdot \left(\frac{\mu}{\rho}\nabla \vec{u}\right) + \vec{f}^{\textrm{int}} + \vec{f}^{\textrm{B}}.
  \label{eq:momentum}
\end{equation}
Here $P$ is the pressure, $\mu$ is the coefficient
of viscosity, $\vec{f}^{\textrm{int}}$ is the interfacial force acting at 
the free surface and at the liquid--solid interface and $\vec{f}^{\textrm{B}}$ is the body force per unit mass acting on the  system. The solid that undergoes melting is assumed to be rigid and interacts with the liquid domain through stresses at the interface. 

The hydrodynamic pressure is $P=p+\tilde{p}$, where $\tilde{p}$ is a background or 
ambient constant pressure which does not contribute to the pressure gradient force due to incompressibility of the medium. 
The pressure $p$ is not coupled to the density and serves to ensure an incompressibility constraint such as a zero divergence velocity field
\begin{equation}
  \nabla \cdot \mathbf{u} = 0,
\label{divv}
\end{equation}
or equivalently, an isochoric deformation given by a unit determinant of the deformation gradient tensor $\vec{F}$ \cite{nair2015volume}
\begin{equation}
  \textrm{det}(\vec{F}) = 1.
\label{defgrad}
\end{equation}

The free surface of the fluid and its intersection with a solid surface 
(called the 
contact line) are subject to capillary forces modelled as inter particle 
forces. 
The numerical implementation of models for heat transfer, phase change, 
capillary fluid flow and the solid-fluid interaction based on Smoothed Particle Hydrodynamics is explained henceforth. 

\section{SPH implementation and validation}
\label{implementation}

The SPH method is based on 
discrete computational nodes that carry field variable values and interact 
with each other within a cut-off radius associated with each node. 
A smoothing function, $W$ (also know as the kernel)
and its derivatives are used to define continuous approximations of the field 
and its derivatives through convolution. Conservation of momentum 
is satisfied in the bulk through interparticle forces obtained through the kernel
approximations.

For an incompressible fluid, the SPH formulation for momentum conservation 
is \cite{szewc2012analysis}:
\begin{equation}
  \begin{split}                                                                 
   \left. \frac{d \vec{u} }{d t}\right| _a = & - \sum_b m_b\frac{p_a + p_b}{\rho_a \rho_b}  \nabla_a W_{ab}  \\
                                  &  + \sum_b m_b  \frac{\mu_a + \mu_b }{\rho_a \rho_b}F_{ab} \vec{u}_{ab} + \vec{f}^{\textrm{int}}_a +\vec{f}_a^B,
      \end{split}  
      \label{eq:sphdiscretization}
\end{equation}
where $m$ is the mass, $\rho$ is the density, $p$ is the hydrodynamic pressure, $\mu$ is the coefficient of viscosity and $\vec{u}$ is the velocity at a particle 
$a$ or particle $b$ in the neighborhood of $a$. Since $\tilde{p}$ is a background 
pressure our implementation assumes $\nabla P = \nabla p$. 
The function $W$ is a radially symmetric and positive definite smoothing function
(also known as the smoothing kernel) with a finite cut off radius 
for the SPH
discretization defined for a particle pair as $W_{ab} = W(r_{ab},h)$, where $h$ is 
the smoothing length of the kernel. The gradient of the smoothing function 
appears in the gradient terms as $\nabla_a W_{ab}$ for a particle $a$ with respect to its neighbor $b$. The radial derivative of the kernel, given by 
$F_{ab}$ \cite{monaghan1992smoothed} is computed from the gradient of $W$ as 
\begin{equation}
  F_{ab}= \frac{\vec{r}_{ab} \cdot \nabla_a W_{ab}}{r_{ab}^{2}+\epsilon^2} 
\end{equation}
where $\epsilon$ is a small number introduced to avoid division by zero in the rare
event of particles overlapping in position and is usually set to $(0.01h)^2$.

The hydrodynamic pressure $p$ in incompressible flows is nothing but a 
Lagrange multiplier that satisfies a constraint for incompressibility given by Eq. \ref{divv} or \ref{defgrad}. For a divergence free velocity field, following 
grid based fluid simulation methods, pressure is obtained by solving the following pressure Poisson equation:
\begin{equation}
  \nabla \cdot \left(\frac{1}{\rho}\nabla p\right) = \frac{\nabla \cdot \vec{u} }{dt}.
  \label{eq:ppe}
\end{equation}
In the discrete SPH domain, the above equation can be approximated as
\begin{equation}
  \sum_{b} \frac{m_b}{\rho_b}\frac{4(p_a- p_b)}{\rho_a +\rho_b }  F_{ab}
     = \sum_{b}\frac{m_b}{\rho_b}  \frac{\vec{u}_{ab} \cdot \nabla_a W_{ab}}{\Delta t} .
\end{equation}
This approximation with unknown pressure values, represents a simultaneous 
system of linear equations in the unknows $p_a$, and can be solved 
numerically using a linear solver such as BiCGSTAB \cite{sleijpen1994bicgstab}. For a domain 
at least partly bounded by free surfaces, a Dirichlet BC
for pressure can be semi-analytically applied by the following modification
to this linear system \cite{nair2014improved}:
 \begin{equation}                                                 
     (p_a - p_o)  \kappa - \sum_{b} \frac{m_b}{\rho_b}\frac{4p_b}{\rho_a +\rho_b} F_{ab}
     = \sum_{b}\frac{m_b}{\rho_b}\left(    \frac{\vec{u}_{ab} \cdot \nabla_a W_{ab}}{\Delta t} - \frac{4p_o}{\rho_a +\rho_b}F_{ab} \right)
               \label{eq:freesurf}                                                            
              \end{equation}
   where $p_0$ represents the ambient pressure and  $\kappa $, given by    
   \begin{equation}                                                             
     \kappa = \sum_{b_{\textrm{bulk}}}\frac{m_b}{\rho_b} \frac{4}{\rho_a +\rho_b}\frac{\vec{r}_{ab} \cdot        
   \nabla_a W_{ab}}{r_{ab}^{2}+\epsilon^2}  ,
     \label{eq:kappa}
   \end{equation}
is a factor which remains constant for a given domain with given smoothing 
paramters and constant density. This modification effectively applies 
a penalty term that accounts for the deficiency of the kernel for a particle
near the free surface, and does not add to the computational cost. 

Capillary forces are modelled using 
an inter-particle force function  $\vec{f}_{ab}^{\textrm{int}}$ based on the molecular theory 
of surface tension \cite{rowlinson2013molecular, tartakovsky2016pairwise} and is chosen to have an attractive part in the long range 
and a repulsive part in the short range. This capillary model is elaborated 
in a recent publication in the context of dynamic free surface flows \cite{nair2018dynamic}. We use the following pairwise force: 
\begin{equation}
  F_{\alpha\beta}^{\textrm{int}} (r_{ab}) =\begin{cases}
    -s_{\alpha\beta}\cos \left( \frac{3\pi}{4} q_{ab} \right) & q_{ab} = \frac{r_{ab}}{h'} \leq q_\textrm{cutoff} \\
    0 & q_{ab} = \frac{r_{ab}}{h'} > q_\textrm{cutoff},
  \end{cases}
  \label{eq:interaction}
\end{equation}
where $\alpha$ and $\beta$ represent the phases of the particles $a$ and $b$, respectively. The strength of the pairwise force is given by $s_{\alpha\beta}$. Here the cut off distance of the pairwise force is set to be the same as that of the smoothing kernel. 
The strength of the 
pairwise force for a given macroscopic surface tension can be predetermined \cite{tartakovsky2016pairwise}
in the presence of a free surface and this relationship is \cite{nair2018dynamic}:
 \begin{align}
   \sigma &= \lambda s_{ll} h_\text{r}^4 , \quad \textrm{for 2D and}  \label{eq:2dsurftens} \\ 
   \sigma &= \lambda s_{ll} \frac{h_\text{r}^5}{\Delta x}  \quad \textrm{for 3D,}
   \label{eq:3dsurftens}
 \end{align}
 respectively.  Here, $h_\text{r}$ is the ratio of the smoothing length 
of the kernel to the initial particle spacing $\Delta x$ (here we use a square lattice arrangement of
particles as the initial condition). Note that these expressions correspond to the specific 
choice of pairwise force function and compact support. The constant due to 
integration of the pairwise force function, $\lambda$ takes the value 
$0.0476$ in two dimensions and $0.0135\pi$ in three dimensions, respectively,
for the interaction function given by eq. \ref{eq:interaction}.  
We use 
an explicit viscous force approximation (the second term on the right hand 
side) that 
is extensively used in SPH literature \cite{morris1997modeling} especially 
for low Reynolds number flows. 

\subsection{Heat transfer across free surface}

Heat transfer problems have been solved in SPH using the approximation of
second order derivatives. In finite domains, the kernel of SPH gets truncated at the interface. Our intended applications require ambient boundary 
conditions. This is achieved by a semi analytic Dirichlet BC for temperature
similar to the application of Dirichlet pressure BC
in the Laplacian of pressure presented in the previous section. 

The heat transfer model (Eq. \ref{eq:heat}) is approximated in the SPH domain as 
follows:
\begin{equation}
  C\rho \frac{\mathrm{d}T}{\mathrm{d}t} = \sum_b \frac{m_b}{\rho_b}\frac{4}{\rho_a+\rho_b}F_{ab}\left(\frac{k_a+k_a}{2}T_{ab}\right)
\label{SPHTemp}
\end{equation}
where, $k_a$ and $k_b$ represent the conductivity at the node $a$ and $b$ 
respectively and $T_{ab} = T_a- T_b$.

Following a similar approach to the free surface Dirichlet BC 
for the pressure Poisson equation \cite{nair2014improved} presented above in Eq. \ref{eq:freesurf}, Dirichlet boundary 
condition for temperature can be applied semi-analytically, as well.
Applying this to the eq. \ref{eq:heat} would lead to the following heat transfer equation
approximation:

\begin{equation}
 \left.  C\rho\frac{\partial T}{\partial t}\right|_a  = \sum_b m_b 2 \frac{k_a + k_b}{\rho_a +\rho_b} F_{ab} \left(T_a - T_b \right) + \kappa\frac{k_a + k^o }{2}\rho_a \left(T_a - T^o \right),
  \label{SPHTemp_fs}
\end{equation}
where $\kappa$ is the same constant as in Eq. \ref{eq:kappa} and $T^o$ is the ambient temperature (which can be time varying). 

The accuracy of heat transfer is central to our problem and requires careful validation. For this, we solve the 2D heat conduction problem for a 
flat plate with free surfaces on all sides and with an adiabatic wall on one side. The adiabatic wall is modeled using static particles.

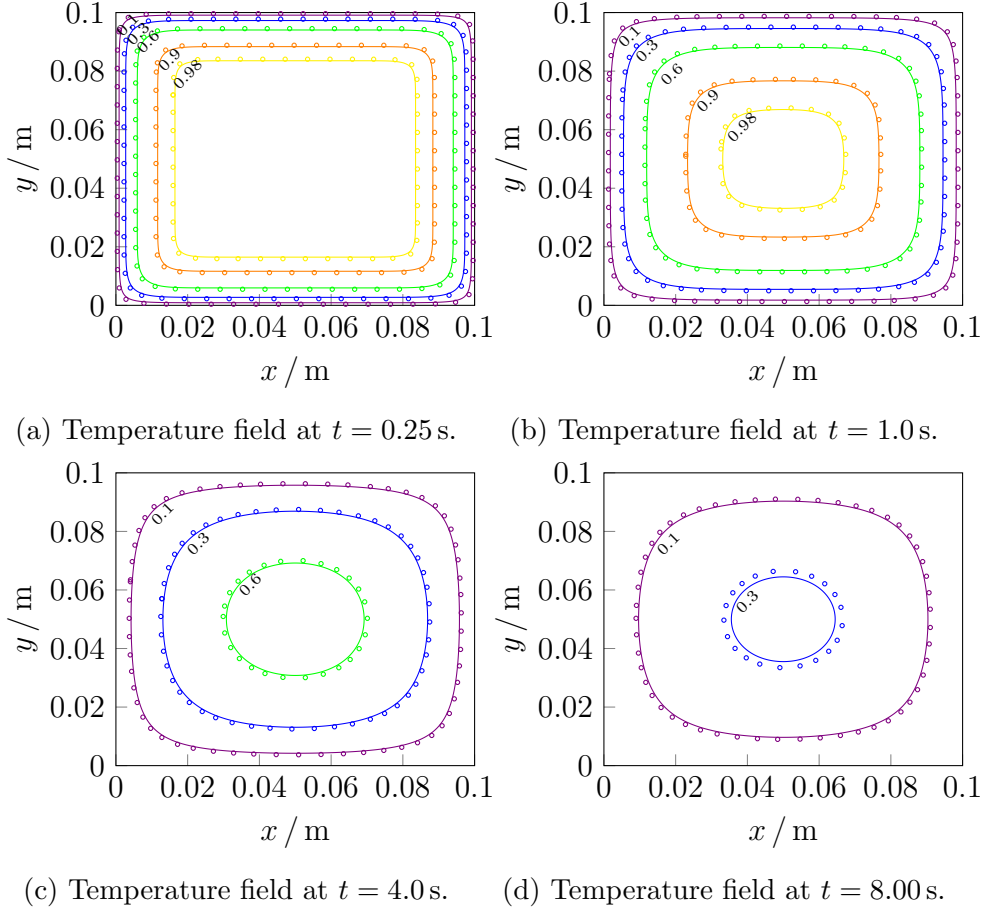
\begin{figure}[!htb]
  \begin{center}
  \begin{subfigure}{2.5in}
		\begin{tikzpicture}
		\begin{axis}[ 
		width=2.5in,
		xlabel = $x\,/\,\mathrm{m}$,
		ylabel = $y\,/\,\mathrm{m}$,
		xtick pos=left,
		ytick pos=left,
		ymin=0.01
		scaled y ticks = false,
		y tick label style={/pgf/number format/fixed,
			/pgf/number format/1000 sep = \thinspace 
		},
		scaled x ticks = false,
		x tick label style={/pgf/number format/fixed,
			/pgf/number format/1000 sep = \thinspace 
		},
		xmin=0,xmax=0.1,
		ymin =0, ymax = 0.1,
		domain=-0.01:0.11, restrict y to domain=-0.01:0.11, restrict x to domain=-0.01:0.11,
		legend pos = north east,
		]
		\addplot[ mark size = 0.8pt,only marks, mark=o, mark repeat=10,draw=violet] table [ col sep=space]{Cont160t0.25_0.1_SPH.txt};
		\addplot[ mark size = 0.8pt, only marks, mark=o, mark repeat=10,draw=blue] table [ col sep=space]{Cont160t0.25_0.3_SPH.txt};
		\addplot[ mark size = 0.8pt, only marks, mark=o, mark repeat=10,draw=green] table [ col sep=space]{Cont160t0.25_0.6_SPH.txt};
		\addplot[ mark size = 0.8pt,only marks, mark=o, mark repeat=10,draw=orange] table [ col sep=space]{Cont160t0.25_0.9_SPH.txt};
		\addplot[ mark size = 0.8pt, only marks, mark=o,draw=yellow, mark repeat=10,] table [ col sep=space]{Cont160t0.25_0.98_SPH.txt};
		\addplot[ mark size = 1.0pt,draw=violet] table [col sep=space, ]{Cont160t0.25_0.1_anl.txt};
		\addplot[ mark size = 1.0pt,draw=blue] table [col sep=space, ]{Cont160t0.25_0.3_anl.txt};
		\addplot[ mark size = 1.0pt,draw=green] table [col sep=space, ]{Cont160t0.25_0.6_anl.txt};
		\addplot[ mark size = 1.0pt,draw=orange] table [col sep=space, ]{Cont160t0.25_0.9_anl.txt};
		\addplot[ mark size = 1.0pt,draw=yellow] table [col sep=space, ]{Cont160t0.25_0.98_anl.txt};
		\pgfplotsset{
			after end axis/.code={
				\node[black,rotate=45] at (axis cs:0.0033,0.096){\tiny{0.1}};
				\node[black,rotate=45] at (axis cs:0.0062,0.0928){\tiny{0.3}};  
				\node[black,rotate=45] at (axis cs:0.009,0.09){\tiny{0.6}};  
				\node[black,rotate=45] at (axis cs:0.015,0.084){\tiny{0.9}};  
				\node[black,rotate=45] at (axis cs:0.02,0.079){\tiny{0.98}};    
			}
		}
		\end{axis}
		\end{tikzpicture}
		\caption{Temperature field at $t=0.25\,\mathrm{s}$.}
	\end{subfigure}
	\begin{subfigure}{2.5in}
		\begin{tikzpicture}
		\begin{axis}[ 
		width=2.5in,
		ymin=0.0,
		xlabel = $x\,/\,\mathrm{m}$,
		ylabel = $y\,/\,\mathrm{m}$,
		xtick pos=left,
		ytick pos=left,
		scaled y ticks = false,
		y tick label style={/pgf/number format/fixed,
			/pgf/number format/1000 sep = \thinspace 
		},
		scaled x ticks = false,
		x tick label style={/pgf/number format/fixed,
			/pgf/number format/1000 sep = \thinspace 
		},
		xmin=0,xmax=0.1,
		ymin =0, ymax = 0.1,
		domain=-0.01:0.11, restrict y to domain=-0.01:0.11, restrict x to domain=-0.01:0.11,
		legend pos = north east,
		]
		\addplot[ mark size = 0.8pt,only marks, mark=o, mark repeat=10,draw=violet] table [ col sep=space]{Cont160t1.0_0.1_SPH.txt};
		\addplot[ mark size = 0.8pt, only marks, mark=o, mark repeat=10,draw=blue] table [ col sep=space]{Cont160t1.0_0.3_SPH.txt};
		\addplot[ mark size = 0.8pt, only marks, mark=o, mark repeat=10,draw=green] table [ col sep=space]{Cont160t1.0_0.6_SPH.txt};
		\addplot[ mark size = 0.8pt,only marks, mark=o, mark repeat=10,draw=orange] table [ col sep=space]{Cont160t1.0_0.9_SPH.txt};
		\addplot[ mark size = 0.8pt, only marks, mark=o,draw=yellow, mark repeat=10,] table [ col sep=space]{Cont160t1.0_0.98_SPH.txt};
		\addplot[ mark size = 1.0pt,draw=violet] table [col sep=space, ]{Cont160t1.0_0.1_anl.txt};
		\addplot[ mark size = 1.0pt,draw=blue] table [col sep=space, ]{Cont160t1.0_0.3_anl.txt};
		\addplot[ mark size = 1.0pt,draw=green] table [col sep=space, ]{Cont160t1.0_0.6_anl.txt};
		\addplot[ mark size = 1.0pt,draw=orange] table [col sep=space, ]{Cont160t1.0_0.9_anl.txt};
		\addplot[ mark size = 1.0pt,draw=yellow] table [col sep=space, ]{Cont160t1.0_0.98_anl.txt};
		\pgfplotsset{
			after end axis/.code={
				\node[black,rotate=45] at (axis cs:0.0072,0.092){\tiny{0.1}};
				\node[black,rotate=45] at (axis cs:0.012,0.087){\tiny{0.3}};  
				\node[black,rotate=45] at (axis cs:0.019,0.079){\tiny{0.6}};  
				\node[black,rotate=45] at (axis cs:0.029,0.07){\tiny{0.9}};  
				\node[black,rotate=45] at (axis cs:0.0385,0.061){\tiny{0.98}};    
			}
		}
		\end{axis}	
		\end{tikzpicture}
		\caption{Temperature field at $t=1.0\,\mathrm{s}$.}
	\end{subfigure}           \\	
	\begin{subfigure}{2.5in}
		\begin{tikzpicture}
		\begin{axis}[ 
		width=2.5in,
		ymin=0.0,
		xlabel = $x\,/\,\mathrm{m}$,
		ylabel = $y\,/\,\mathrm{m}$,
		xtick pos=left,
		ytick pos=left,
		scaled y ticks = false,
		y tick label style={/pgf/number format/fixed,
			/pgf/number format/1000 sep = \thinspace 
		},
		scaled x ticks = false,
		x tick label style={/pgf/number format/fixed,
			/pgf/number format/1000 sep = \thinspace 
		},
		xmin=0,xmax=0.1,
		ymin =0, ymax = 0.1,
		domain=-0.01:0.11, restrict y to domain=-0.01:0.11, restrict x to domain=-0.01:0.11,
		legend pos = north east,
		]
		\addplot[ mark size = 0.8pt,only marks, mark=o, mark repeat=10,draw=violet] table [ col sep=space]{Cont160t4.0_0.1_SPH.txt};
		\addplot[ mark size = 0.8pt, only marks, mark=o, mark repeat=10,draw=blue] table [ col sep=space]{Cont160t4.0_0.3_SPH.txt};
		\addplot[ mark size = 0.8pt, only marks, mark=o, mark repeat=10,draw=green] table [ col sep=space]{Cont160t4.0_0.6_SPH.txt};
		
		\addplot[ mark size = 1.0pt,draw=violet] table [col sep=space, ]{Cont160t4.0_0.1_anl.txt};
		\addplot[ mark size = 1.0pt,draw=blue] table [col sep=space, ]{Cont160t4.0_0.3_anl.txt};
		\addplot[ mark size = 1.0pt,draw=green] table [col sep=space, ]{Cont160t4.0_0.6_anl.txt};
		\pgfplotsset{
			after end axis/.code={
				\node[black,rotate=45] at (axis cs:0.013,0.086){\tiny{0.1}};
				\node[black,rotate=45] at (axis cs:0.023,0.076){\tiny{0.3}};  
				\node[black,rotate=45] at (axis cs:0.0375,0.062){\tiny{0.6}};  
			}
		}
		\end{axis}
		
		\end{tikzpicture}
		\caption{Temperature field at $t=4.0\,\mathrm{s}$.}
	\end{subfigure}
	\begin{subfigure}{2.5in}
		\begin{tikzpicture}
		\begin{axis}[ 
		width=2.5in,
		ymin=0.0,
		xlabel = $x\,/\,\mathrm{m}$,
		ylabel = $y\,/\,\mathrm{m}$,
		xtick pos=left,
		ytick pos=left,
		scaled y ticks = false,
		y tick label style={/pgf/number format/fixed,
			/pgf/number format/1000 sep = \thinspace 
		},
		scaled x ticks = false,
		x tick label style={/pgf/number format/fixed,
			/pgf/number format/1000 sep = \thinspace 
		},
		xmin=0,xmax=0.1,
		ymin =0, ymax = 0.1,
		domain=-0.01:0.11, restrict y to domain=-0.01:0.11, restrict x to domain=-0.01:0.11,
		legend pos = north east,
		]
		\addplot[ mark size = 0.8pt,only marks, mark=o, mark repeat=10,draw=violet] table [ col sep=space]{Cont160t8.0_0.1_SPH.txt};
		\addplot[ mark size = 0.8pt, only marks, mark=o, mark repeat=10,draw=blue] table [ col sep=space]{Cont160t8.0_0.3_SPH.txt};
		
		\addplot[ mark size = 1.0pt,draw=violet] table [col sep=space, ]{Cont160t8.0_0.1_anl.txt};
		\addplot[ mark size = 1.0pt,draw=blue] table [col sep=space, ]{Cont160t8.0_0.3_anl.txt};
		\pgfplotsset{
			after end axis/.code={
				\node[black,rotate=45] at (axis cs:0.018,0.076){\tiny{0.1}};
				\node[black,rotate=45] at (axis cs:0.04,0.056){\tiny{0.3}};  
			}
		}
		\end{axis}
		
		\end{tikzpicture}
		\caption{Temperature field at $t=8.00\,\mathrm{s}$.}
		\label{Val:Heat4Dd}
	\end{subfigure}
	\caption{Evolution of temperature profile within a square at different times 
  for a resolution of $160\times160$ particles with Dirichlet BC
  on all sides.}
	\label{Val:Heat4D}
\end{center}
\end{figure}

\begin{figure}[!htb]
  \begin{center}
	\begin{subfigure}{2.5in}
		\begin{tikzpicture}
		\begin{axis}[ 
		width=2.5in,
		ymin=0.0,
		xlabel = $x\,/\,\mathrm{m}$,
		ylabel = $y\,/\,\mathrm{m}$,
		xtick pos=left,
		ytick pos=left,
		scaled y ticks = false,
		y tick label style={/pgf/number format/fixed,
			/pgf/number format/1000 sep = \thinspace 
		},
		scaled x ticks = false,
		x tick label style={/pgf/number format/fixed,
			/pgf/number format/1000 sep = \thinspace 
		},
		xmin=0,xmax=0.1,
		ymin =0, ymax = 0.1,
		domain=-0.01:0.11, restrict y to domain=-0.01:0.11, restrict x to domain=-0.01:0.11,
		legend pos = north east,
		]
		\addplot[ mark size = 0.8pt,only marks, mark=o, mark repeat=10,draw=violet] table [ col sep=space]{Cont160t0.25_0.1_3DN1_SPH.txt};
		\addplot[ mark size = 0.8pt, only marks, mark=o, mark repeat=10,draw=blue] table [ col sep=space]{Cont160t0.25_0.3_3DN1_SPH.txt};
		\addplot[ mark size = 0.8pt, only marks, mark=o, mark repeat=10,draw=green] table [ col sep=space]{Cont160t0.25_0.6_3DN1_SPH.txt};
		\addplot[ mark size = 0.8pt,only marks, mark=o, mark repeat=10,draw=orange] table [ col sep=space]{Cont160t0.25_0.9_3DN1_SPH.txt};
		\addplot[ mark size = 0.8pt, only marks, mark=o,draw=yellow, mark repeat=10,] table [ col sep=space]{Cont160t0.25_0.98_3DN1_SPH.txt};
		
		\addplot[ mark size = 1.0pt,draw=violet] table [col sep=space, ]{Cont160t0.25_0.1_3DN1_anl.txt};
		\addplot[ mark size = 1.0pt,draw=blue] table [col sep=space, ]{Cont160t0.25_0.3_3DN1_anl.txt};
		\addplot[ mark size = 1.0pt,draw=green] table [col sep=space, ]{Cont160t0.25_0.6_3DN1_anl.txt};
		\addplot[ mark size = 1.0pt,draw=orange] table [col sep=space, ]{Cont160t0.25_0.9_3DN1_anl.txt};
		\addplot[ mark size = 1.0pt,draw=yellow] table [col sep=space, ]{Cont160t0.25_0.98_3DN1_anl.txt};
		\pgfplotsset{
			after end axis/.code={
				\node[black,rotate=0] at (axis cs:0.0033,0.096){\tiny{0.1}};
				\node[black,rotate=0] at (axis cs:0.0062,0.09){\tiny{0.3}};  
				\node[black,rotate=0] at (axis cs:0.009,0.084){\tiny{0.6}};  
				\node[black,rotate=0] at (axis cs:0.015,0.076){\tiny{0.9}};  
				\node[black,rotate=0] at (axis cs:0.021,0.07){\tiny{0.98}};    
        \node[black,rotate=0] at (axis cs:0.051,0.093){\small{\textsf{Adiabatic}}};    
        \node[black,rotate=0] at (axis cs:0.051,0.085){\small{ \textsf{wall}}};    
			}
		}
		\end{axis}
		
		\end{tikzpicture}
		\caption{Temperature field at  $t=0.25\,\mathrm{s}$.}
	\end{subfigure}
	\begin{subfigure}{2.5in}
		\begin{tikzpicture}
		\begin{axis}[ 
		width=2.5in,
		ymin=0.0,
		xlabel = $x\,/\,\mathrm{m}$,
		ylabel = $y\,/\,\mathrm{m}$,
		xtick pos=left,
		ytick pos=left,
		scaled y ticks = false,
		y tick label style={/pgf/number format/fixed,
			/pgf/number format/1000 sep = \thinspace 
		},
		scaled x ticks = false,
		x tick label style={/pgf/number format/fixed,
			/pgf/number format/1000 sep = \thinspace 
		},
		xmin=0,xmax=0.1,
		ymin =0, ymax = 0.1,
		domain=-0.01:0.11, restrict y to domain=-0.01:0.11, restrict x to domain=-0.01:0.11,
		legend pos = north east,
		]
		\addplot[ mark size = 0.8pt,only marks, mark=o, mark repeat=10,draw=violet] table [ col sep=space]{Cont160t1.0_0.1_3DN1_SPH.txt};
		\addplot[ mark size = 0.8pt, only marks, mark=o, mark repeat=10,draw=blue] table [ col sep=space]{Cont160t1.0_0.3_3DN1_SPH.txt};
		\addplot[ mark size = 0.8pt, only marks, mark=o, mark repeat=10,draw=green] table [ col sep=space]{Cont160t1.0_0.6_3DN1_SPH.txt};
		\addplot[ mark size = 0.8pt,only marks, mark=o, mark repeat=10,draw=orange] table [ col sep=space]{Cont160t1.0_0.9_3DN1_SPH.txt};
		\addplot[ mark size = 0.8pt, only marks, mark=o,draw=yellow, mark repeat=10,] table [ col sep=space]{Cont160t1.0_0.98_3DN1_SPH.txt};
		
		\addplot[ mark size = 1.0pt,draw=violet] table [col sep=space, ]{Cont160t1.0_0.1_3DN1_anl.txt};
		\addplot[ mark size = 1.0pt,draw=blue] table [col sep=space, ]{Cont160t1.0_0.3_3DN1_anl.txt};
		\addplot[ mark size = 1.0pt,draw=green] table [col sep=space, ]{Cont160t1.0_0.6_3DN1_anl.txt};
		\addplot[ mark size = 1.0pt,draw=orange] table [col sep=space, ]{Cont160t1.0_0.9_3DN1_anl.txt};
		\addplot[ mark size = 1.0pt,draw=yellow] table [col sep=space, ]{Cont160t1.0_0.98_3DN1_anl.txt};
		\pgfplotsset{
			after end axis/.code={
				\node[black,rotate=0] at (axis cs:0.051,0.0035){\tiny{0.1}};
				\node[black,rotate=0] at (axis cs:0.051,0.007){\tiny{0.3}};  
				\node[black,rotate=0] at (axis cs:0.051,0.014){\tiny{0.6}};  
				\node[black,rotate=0] at (axis cs:0.051,0.0255){\tiny{0.9}};  
				\node[black,rotate=0] at (axis cs:0.051,0.036){\tiny{0.98}};    
			}
		}
		\end{axis}
		
		\end{tikzpicture}
		\caption{Temperature field at  $t=1.00\,\mathrm{s}$.}
	\end{subfigure}\\
	\begin{subfigure}{2.5in}
		\begin{tikzpicture}
		\begin{axis}[ 
		width=2.5in,
		ymin=0.0,
		xlabel = $x\,/\,\mathrm{m}$,
		ylabel = $y\,/\,\mathrm{m}$,
		xtick pos=left,
		ytick pos=left,
		scaled y ticks = false,
		y tick label style={/pgf/number format/fixed,
			/pgf/number format/1000 sep = \thinspace 
		},
		scaled x ticks = false,
		x tick label style={/pgf/number format/fixed,
			/pgf/number format/1000 sep = \thinspace 
		},
		xmin=0,xmax=0.1,
		ymin =0, ymax = 0.1,
		domain=-0.01:0.11, restrict y to domain=-0.01:0.11, restrict x to domain=-0.01:0.11,
		legend pos = north east,
		]
		\addplot[ mark size = 0.8pt,only marks, mark=o, mark repeat=10,draw=violet] table [ col sep=space]{Cont160t4.0_0.1_3DN1_SPH.txt};
		\addplot[ mark size = 0.8pt, only marks, mark=o, mark repeat=10,draw=blue] table [ col sep=space]{Cont160t4.0_0.3_3DN1_SPH.txt};
		\addplot[ mark size = 0.8pt, only marks, mark=o, mark repeat=10,draw=green] table [ col sep=space]{Cont160t4.0_0.6_3DN1_SPH.txt};
		
		\addplot[ mark size = 1.0pt,draw=violet] table [col sep=space, ]{Cont160t4.0_0.1_3DN1_anl.txt};
		\addplot[ mark size = 1.0pt,draw=blue] table [col sep=space, ]{Cont160t4.0_0.3_3DN1_anl.txt};
		\addplot[ mark size = 1.0pt,draw=green] table [col sep=space, ]{Cont160t4.0_0.6_3DN1_anl.txt};
		\pgfplotsset{
			after end axis/.code={
				\node[black,rotate=0] at (axis cs:0.05,0.032){\tiny{0.1}};
				\node[black,rotate=0] at (axis cs:0.05,0.016){\tiny{0.3}};  
				\node[black,rotate=0] at (axis cs:0.05,0.007){\tiny{0.6}};  
			}
		}
		\end{axis}
		
		\end{tikzpicture}
		\caption{Temperature field at $t=4.00\,\mathrm{s}$.}
	\end{subfigure}
	\begin{subfigure}{2.5in}
		\begin{tikzpicture}
		\begin{axis}[ 
		width=2.5in,
		ymin=0.0,
		xlabel = $x\,/\,\mathrm{m}$,
		ylabel = $y\,/\,\mathrm{m}$,
		xtick pos=left,
		ytick pos=left,
		scaled y ticks = false,
		y tick label style={/pgf/number format/fixed,
			/pgf/number format/1000 sep = \thinspace 
		},
		scaled x ticks = false,
		x tick label style={/pgf/number format/fixed,
			/pgf/number format/1000 sep = \thinspace 
		},
		xmin=0,xmax=0.1,
		ymin =0, ymax = 0.1,
		domain=-0.01:0.11, restrict y to domain=-0.01:0.11, restrict x to domain=-0.01:0.11,
		legend pos = north east,
		]
		\addplot[ mark size = 0.8pt,only marks, mark=o, mark repeat=10,draw=violet] table [ col sep=space]{Cont160t8.0_0.1_3DN1_SPH.txt};
		\addplot[ mark size = 0.8pt, only marks, mark=o, mark repeat=10,draw=blue] table [ col sep=space]{Cont160t8.0_0.3_3DN1_SPH.txt};
		
		\addplot[ mark size = 1.0pt,draw=violet] table [col sep=space, ]{Cont160t8.0_0.1_3DN1_anl.txt};
		\addplot[ mark size = 1.0pt,draw=blue] table [col sep=space, ]{Cont160t8.0_0.3_3DN1_anl.txt};
		\pgfplotsset{
			after end axis/.code={
				\node[black,rotate=0] at (axis cs:0.05,0.032){\tiny{0.3}};
				\node[black,rotate=0] at (axis cs:0.05,0.012){\tiny{0.1}};  
			}
		}
		\end{axis}
		
		\end{tikzpicture}
		\caption{Temperature field at $t=8.00\,\mathrm{s}$.}
	\end{subfigure}
	\caption{Evolution of temperature profile within a square at different times 
  for a resolution of $160\times160$ particles with Dirichlet BC on three sides 
  and 
  Neumann on the fourth.}
	\label{fig3DN1Hges}
\end{center}
\end{figure}
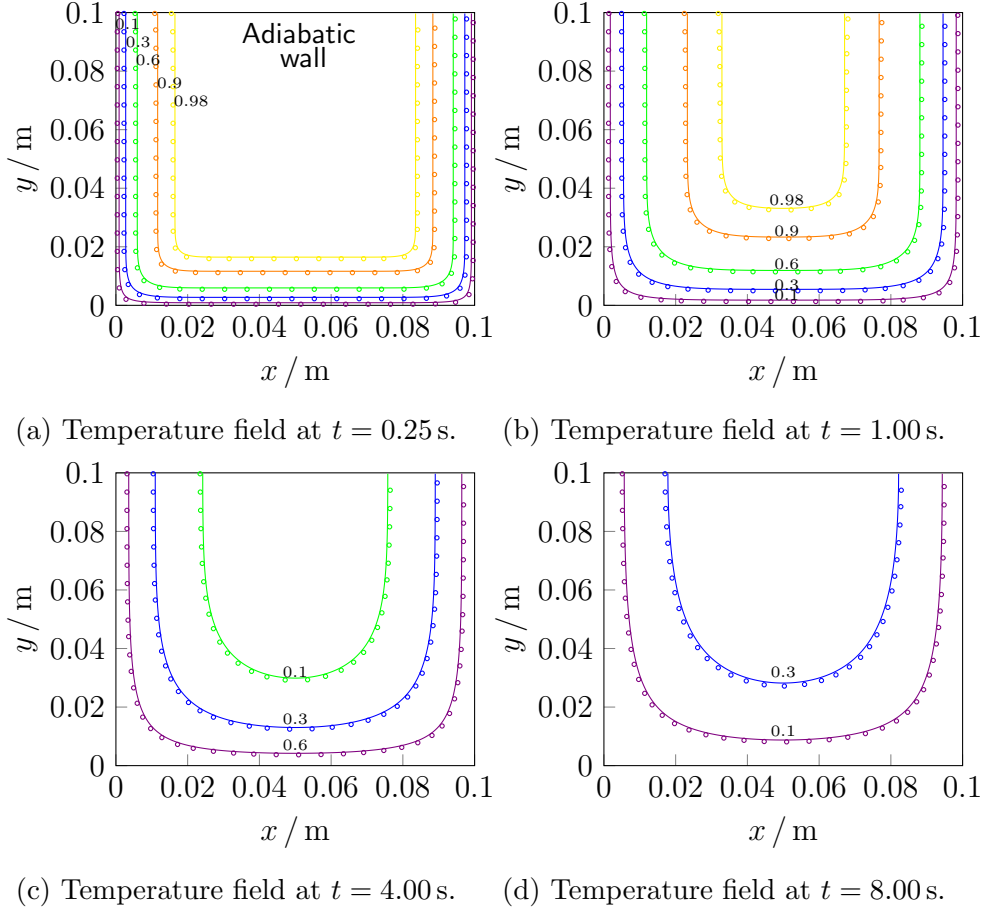
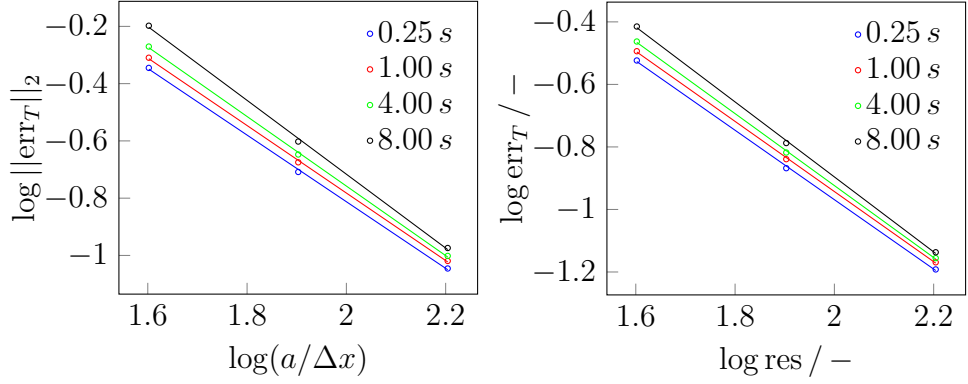
\begin{figure}[!htb]
  \begin{center}
	\begin{subfigure}{2.5in}
		\begin{tikzpicture}
		\begin{axis}[ 
		width=2.5in,
      xlabel = $\log (a/\Delta x)$,
		ylabel = $\log ||\mathrm{err}_T||_2 $,
		xtick pos=left,
		ytick pos=left,
		legend pos = north east,
		legend style={draw=none}
		]
		\addplot+[ mark size = 1.0pt, only marks, draw=blue, mark=o] table [x index = 0, y index = 1 ]{HeatErr4D.txt};
			\addplot+[ mark size = 1.0pt, only marks, draw=red, mark=o] table [col sep=space, x index = 0, y index = 2 ]{HeatErr4D.txt};
			\addplot+[ mark size = 1.0pt, only marks, draw=green, mark=o] table [col sep=space, x index = 0, y index = 3 ]{HeatErr4D.txt};
			\addplot+[ mark size = 1.0pt, only marks, draw=black, mark=o] table [col sep=space, x index = 0, y index = 4 ]{HeatErr4D.txt};
			\addplot[mark size = 1.0pt, draw=blue, domain=1.60206:2.20412] {-1.1636*x+1.5148};
			\addplot[mark size = 1.0pt, draw=red, domain=1.60206:2.20412] {-1.1789*x+1.5757};
			\addplot[mark size = 1.0pt, draw=green, domain=1.60206:2.20412] {-1.2145*x+1.6713};
			\addplot[mark size = 1.0pt, draw=black, domain=1.60206:2.20412] {-1.2895*x+1.8627};
		
		\addlegendentry{$0.25\,s$};
		\addlegendentry{$1.00\,s$};
		\addlegendentry{$4.00\,s$};
		\addlegendentry{$8.00\,s$};
		\end{axis}
		\end{tikzpicture}
		\caption{Dirichlet BC on all sides \newline}	
	\end{subfigure}	
	\begin{subfigure}{2.5in}
		\begin{tikzpicture}
		\begin{axis}[ 
		width=2.5in,
		xlabel = $\log \mathrm{res}\,/\,-$,
		ylabel = $\log \mathrm{err}_T\,/\,- $,
		xtick pos=left,
		ytick pos=left,
		legend pos = north east,
		legend style={draw=none}
		]
			\addplot+[ mark size = 1.0pt, only marks, draw=blue, mark=o] table [col sep=space, x index=0, y index=1 ]{HeatErr3DN1.txt};
			\addplot+[ mark size = 1.0pt, only marks, draw=red, mark=o] table [col sep=space, x index=0, y index=2 ]{HeatErr3DN1.txt};
			\addplot+[ mark size = 1.0pt, only marks, draw=green, mark=o] table [col sep=space, x index=0, y index=3 ]{HeatErr3DN1.txt};
			\addplot+[ mark size = 1.0pt, only marks, draw=black, mark=o] table [col sep=space, x index=0, y index=4 ]{HeatErr3DN1.txt};
			
			\addplot[mark size = 1.0pt, draw=blue, domain=1.60206:2.20412] {-1.1101*x+1.2515};
			\addplot[mark size = 1.0pt, draw=red, domain=1.60206:2.20412] {-1.1216*x+1.3005};
			\addplot[mark size = 1.0pt, draw=green, domain=1.60206:2.20412] {-1.1508*x+1.378};
			\addplot[mark size = 1.0pt, draw=black, domain=1.60206:2.20412] {-1.1993*x+1.5027};
		\addlegendentry{$0.25\,s$};
		\addlegendentry{$1.00\,s$};
		\addlegendentry{$4.00\,s$};
		\addlegendentry{$8.00\,s$};
		\end{axis}
		\end{tikzpicture}
		\caption{Dirichlet BC on three sides and homogenous Neumann BC on one. }
	\end{subfigure}
  \caption{$L_2$ norm of the error in temperature ($T$) in the domain at 
  different times for different
  spatial resolution. The plot shows the order of accuracy is greater than 
  1 at all time instances.}
	\label{fig:l2norm}
\end{center}
\end{figure}
The heat conduction through the flat plate across the free surface is compared 
to the analytical solution at different times in Fig. \ref{Val:Heat4D} where the 
plate is heated at constant temperature from the ambience on all four sides. An array of 
$80\times80$ fixed particles are used to represent a square flat plate of side $0.1$ m. The 
ambient temperature is set to $0$ K and the plate is set to a uniform intial
temperature of $1 $ K. The density of the material is set to $1$ kgm$^-3$ and the 
thermal diffusivity is set to $1\times 10^{-4}$ m$^2$s$-1$.

In  Fig. \ref{fig3DN1Hges},
one of the walls of the plate is set to be adiabatic wall where a homogeneous 
Neumann BC is applied for temperature by 
setting zero temperature gradient  
between particles of the domain and the adiabatic wall. The solutions 
are compared to analytical solutions, available widely in standard 
text books on numerical methods \cite{kreyszig2010advanced}, and a very good match is observed. 

The consistency of this heat transfer model can be seen in Fig. \ref{fig:l2norm}, 
showing greater than first order accurate. The $L_2$ norm of the error at different times
are presented against the resolution of the domain. The order of accuracy of the heat 
transfer simulation with the free surface BC 
can be clearly seen to me more than first order in Fig. \ref{fig:l2norm}. 
\subsection{Static phase change }
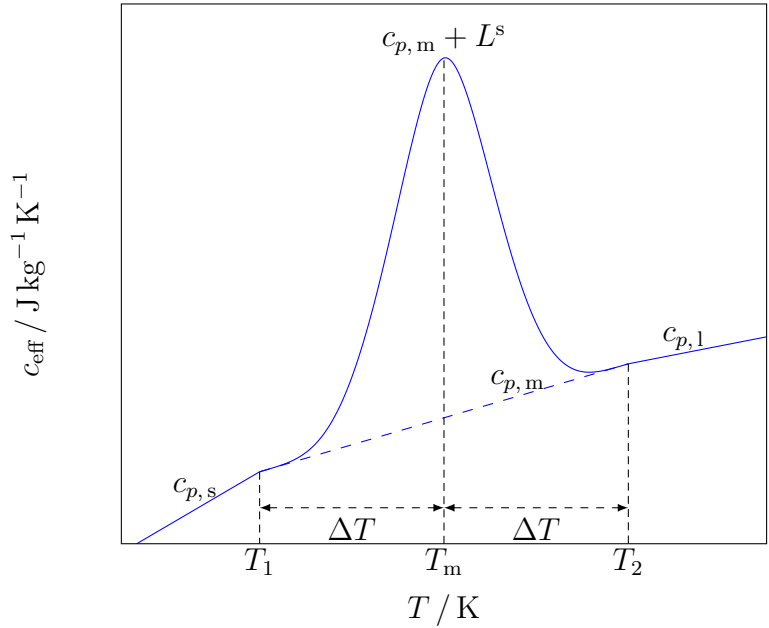
\begin{figure}[!htb]
  \begin{center}
	\begin{tikzpicture}
	\begin{axis}[ 
	width=4in,
	xmin = -3.5, xmax =3.5,
	ymin=-0.2, ymax=1.3,
	xticklabels={,,},
	yticklabels={,,},
	ticks=none,
	xlabel = $T\,/\,\mathrm{K}$,
	ylabel = $c_{\text{eff}}\,/\,\mathrm{J}\,\mathrm{kg}^{-1}\,\mathrm{K}^{-1}$,
	legend pos = north east,
	legend style={draw=none}
	]
	\addplot[mark size = 1.0pt, draw=blue, domain=2:3.5] {0.05*x+0.2};
	\addplot[mark size = 1.0pt, draw=blue, domain=-3.5:-2] {0.15*x+0.3};
    \addplot[mark size = 1.0pt, draw=blue, domain=-2:2, samples=200] {0.075*x+0.15+((1-abs(x)/2)^4)*(2*abs(x)+1) };
    \addplot[mark size = 1.0pt, draw=blue, dash pattern={on 4pt off 4pt}, domain=-2:2, samples=200] {0.075*x+0.15 };
	
	\addplot +[mark=none, draw=black, dash pattern={on 3pt off 2pt}] coordinates {(-2, 0) (-2, -0.2)};
	\addplot +[mark=none, draw=black, dash pattern={on 3pt off 2pt}] coordinates {(2, 0.3) (2, -0.2)};
	\addplot +[mark=none, draw=black, dash pattern={on 3pt off 2pt}] coordinates {(0, -0.2) (0, 1.15)};
	
	\pgfplotsset{
		after end axis/.code={
			\node[black] at (axis cs:-2.7,-0.05){$c_{p,\,\mathrm{s}}$};    			
			\node[black] at (axis cs:2.6,0.37){$c_{p,\,\mathrm{l}}$};
			\node[black] at (axis cs:0,1.2){$c_{p,\,\mathrm{m}}+L^{\mathrm{s}}$};
			\node[black] at (axis cs:0.8,0.25){$c_{p,\,\mathrm{m}}$};
			\draw[latex-latex, dashed] (axis cs: 0,-0.1)--(axis cs: 2,-0.1) node[pos=0.5,below]{$\Delta T$};    
			\draw[latex-latex, dashed] (axis cs: -2,-0.1)--(axis cs: 0,-0.1) node[pos=0.5,below]{$\Delta T$}; 
			
			\node[black] at (axis cs:-2.0,-0.25){$T_{1}$};	
			\node[black] at (axis cs:0,-0.25){$T_{\mathrm{m}}$};
			\node[black] at (axis cs:2.0,-0.25){$T_{2}$};			
		}
	}
	\end{axis}
	\end{tikzpicture}
	\caption{Smoothed effective heat capacity within the mushy region of the phase transition temperature.}
	\label{SPHfigL}
  \end{center}
  \end{figure}

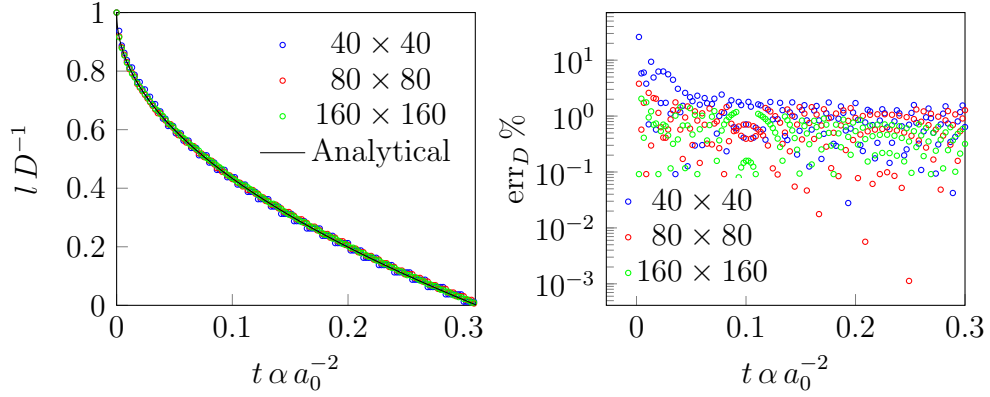
\begin{figure}[!htb]
  \begin{center}
	\begin{subfigure}[t]{2.5in}	
		\begin{tikzpicture}
		\begin{axis}[ 
		width=2.5in,
		xmin = 0, xmax =0.31,
		ymin=0.0, ymax=1,
		ylabel = $l\,D^{-1}$,
		xlabel = $t\,\alpha\,a_0^{-2}  $,
		xtick pos=left,
		ytick pos=left,
		legend pos = north east,
		legend style={draw=none},
		clip mode=individual
		]
		
		\addplot[ mark size = 1.0pt, only marks, draw=blue, mark=o] table [col sep=space, x index = 0, y index = 1 ]{80x80sph.txt};
		\addplot[ mark size = 1.0pt, only marks, draw=red, mark=o] table [col sep=space, x index = 0, y index = 1 ]{160x160sph.txt};
		\addplot[ mark size = 1.0pt, only marks, draw=green, mark=o] table [col sep=space, x index = 0, y index = 1 ]{320x320sph.txt};
		index = 1 ]{80x80anl.txt};
		\addplot[ mark size = 1.0pt, draw=black, mark repeat=4] table [col sep=space, x index = 0, y index = 1 ]{80x80anl.txt};	
		\addlegendentry{$40\times40$};
		\addlegendentry{$80\times80$};
		\addlegendentry{$160\times160$};
		\addlegendentry{Analytical};
		\end{axis}	
		\end{tikzpicture}	
		\caption{Dimensionless interface position along diagonal vs dimensionless time.}
		\label{interface2D}
	\end{subfigure}
	\begin{subfigure}[t]{2.5in}
		\begin{tikzpicture}
		\begin{axis}[ 
		width=2.5in,
		xlabel = $t\,\alpha\,a_0^{-2}  $,
    ylabel = $\textrm{err}_{D}\,\% $ ,
		ymode=log,
    xmax =0.3,
		xtick pos=left,
		ytick pos=left,
		legend pos = south west,
		legend style={draw=none}
		]
      \addplot[ mark size = 1.0pt, only marks, draw=blue,mark=o] table [col sep=space, x index = 0, y index = 2 ]{80x80sph.txt};
  		\addplot+[ mark size = 1.0pt, only marks, draw=red,mark=o] table [col sep=space, x index = 0, y index = 2 , mark repeat=1]{160x160sph.txt};
	  	\addplot[ mark size = 1.0pt, only marks, draw=green, mark =o] table [col sep=space, x index = 0, y index = 2 ]{320x320sph.txt};

		\addlegendentry{$40\times40$};
		\addlegendentry{$80\times80$};
		\addlegendentry{$160\times160$};
		
		\end{axis}
		\end{tikzpicture}
    \caption{Error in interface position}
	\label{interface2D_error}
	\end{subfigure}
	\caption{Stefan's problem: Interface position (a) and calculated errors (b) obtained for different resolutions. }
	\label{interface}
\end{center}
\end{figure}

\begin{figure}[!htb]
  \begin{center}
    \includegraphics[width=5in]{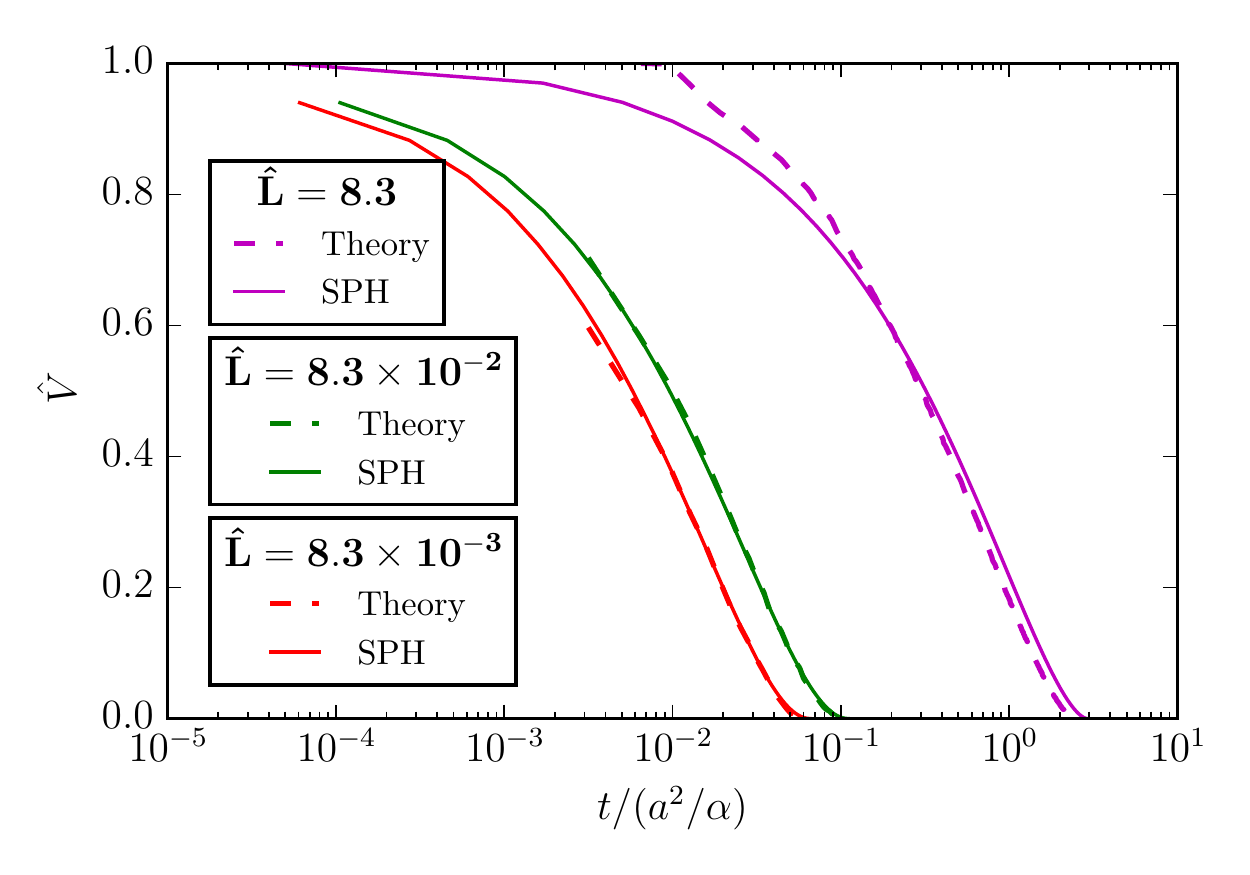}
    \caption{Static melting of a spherical solid--comparison of evolution of solid volume fraction with 1-D axisymmetric theory }
    \label{staticmelting3D}
  \end{center}
\end{figure}

Phase change is modelled using the effective heat capacity $C$ defined in Eq. \ref{eq:effectiveheatcapacity}. There are different strategies for implementing the effect of latent heat into a numerical method. On the one hand it can be added as a source term to the heat conduction equation \cite{Voller1987,Pasandideh-Fard1996} or, on the other hand, by modifying the heat capacity itself \cite{farrokhpanah2017new, Thomas1984,Hsiao1986}. 
In this work an integral interpolation introduced by \cite{farrokhpanah2017new} is implemented to modify the heat capacity in order to model phase change processes because of its low computational costs.
Therefore, the model for a smoothed latent heat in {SPH} can be derived by starting from the definition of an effective heat capacity $C_{\text{eff}}$ which includes the effect of latent heat \cite{Bonacina1973, Hsiao1983} given by Eq. \ref{eq:effectiveheatcapacity}. This formulation is composed of the heat capacities for solid $C_{\mathrm{s}}$ and $C_{\mathrm{l}}$ liquid phase as well as the heat capacity within the phase change region $C_{\mathrm{m}}$. Here, $C_{\mathrm{eff}}$ at the melting temperature $T_m$ includes the total latent heat of the phase change, $L  = \int_{-\infty}^{\infty}L\delta\left(T-T_m\right)$.

The Dirac function can be replaced in the one-dimensional temperature domain by a kernel function $ W_{T} $ with smoothing length $h_T$ analogous to the spatial kernel function in {SPH} resulting in:
\begin{equation}
\label{LH2}
  {C_\textrm{eff}}= \begin{cases}
C_{\mathrm{s}} & T<T_m-\Delta T\\
C_{\mathrm{m}}+LW_{T}\left(T-T_{\mathrm{m}}, h_{\mathrm{T}}\right)&T_{\mathrm{m}}-\Delta T\leq T \leq T_{\mathrm{m}}+\Delta T\\
C_{\mathrm{l}} & T > T_{\mathrm{m}}+\Delta T\\
\end{cases}
\end{equation}
The temperature domain $\Delta T$ represents a mushy region around the phase change temperature. The size of this region is defined as the product of the smoothing length $h_{\mathrm{T}}$ and the maximum range of support $q_{\text{max}}$ of the temperature kernel.
\begin{equation}
\Delta T = h_{T}q_{\mathrm{max}}.
\end{equation}

The outcome of this is a changing $C_{\mathrm{eff}}$ inside of this region with a maximum latent heat effect at the transition temperature. The accuracy of this approach is further improved by considering the distance between adjacent particles and the temperature difference with respect to the transition temperature.
This results in a smoothed latent heat $L^{\mathrm{s}}$ \cite{farrokhpanah2017new} for a particle $a$ defined as
\begin{equation}
L_a^s = \sum_{b} \frac{m_b}{\rho_b}\left(L W_{T}\left(T_b-T_{\mathrm{m}},  h_T\right)\right)W\left(\boldsymbol{r_b}-\boldsymbol{r_a}, h\right).
\end{equation}

This results in the effective heat capacity given in equation (\ref{CeffF}) which is replacing the specific heat capacity of the heat conduction equation (\ref{SPHTemp}).
\begin{equation}
\label{CeffF}
C_{\mathrm{eff}} = \begin{cases}
C_{\mathrm{s}} & T<T_{\mathrm{m}}-\Delta T\\
C_{\mathrm{m}}+L^{\mathrm{s}}&T_{\mathrm{m}}-\Delta T\leq T \leq T_{\mathrm{m}}+\Delta T\\
C_{\mathrm{l}} & T > T_{\mathrm{m}}+\Delta T\\
\end{cases}
\end{equation}

The apparent heat capacity with a smoothed jump for the latent heat is shown in  Fig. \ref{SPHfigL}.

The phase change model in SPH together with the free surface BC for ambient temperature needs to be validated. Here we choose the classical Stefan's problem in 2D and an axisymmetric version in 3D to validate the evolution of the melting interface with time.

In two dimensions, the two edges of a right angled corner of a 2D square plate is heated. The medium is set to a density of 1 kg$/$m$^3$. The initial and melting temperature are set to $2$ and $2.3$ K respectively. The conductivity and specific heat capacity of both phases were maintained at a value of $1$ in SI units. The latent heat capacity was set to $0.25$ J$/$kg. Three different resolutions were considered corresponding to 40, 80 and 160 particles along one direction in the quarter space of dimensions $a=3$ m. The choice of these parameters are arbitrary, since the aim of the validation is only to check the numerical approximation of the model. In the following section we will resort to more realistic simulation parameters. As the temperature rises above the melting temperature, the solid melts and the solid liquid interface moves inwards. The location of the phase change interface along the diagonal, normalized by the width of 
the plate is plotted against non-dimensional time in Fig. \ref{interface2D}.
The melting process is assumed to be static, such that the molten region is assumed to not deform. This problem is simulated, without liquid deformation for different spatial resolution and compared against the analytical solution. Note that the heat transfer across the free boundary is based on the semi-analytic boundary formulation in Eq. \ref{eq:freesurf}. The interface location appears as a stepped curve because of the discrete sampling in time. We see that the interface evolution is predicted accurately. In Fig. \ref{interface2D_error} we see the order of accuracy increase with increasing resolution, consistently. The discrete sampling in time makes it difficult to obtain the exact order of accuracy.

In three dimensions, we solve the Stefan's problem in a sphere and compare the results with the
solutions of 
a one dimensional axisymmetric model proposed in \cite{mccue2008classical}, given by
\begin{equation}
  \frac{\partial h }{\partial \hat{t} } = \frac{\partial^2 \hat{T}}{\partial r^2} + \frac{2}{r}\frac{\partial \hat{T}}{\partial r} \quad \text{in}\quad 0<r<1,
  \label{theory3D}
\end{equation}
where $r$ is the radial location in the unit sphere, normalized by the radius of 
the sphere, R. Time is normalized by $R^2/k_s$, where $k_s$ is the conductivity of 
the solid phase. The enthalpy $h$ is related to the temperature $\hat{T}$ (normalized by $\Delta t  =T_a - T_m$) by 
\begin{equation}
  T = \begin{cases}  h-\beta, & h< 0 \\ 0, & 0\leq h \leq \beta , \\ \tilde{k} (h-\beta), & h>\beta, \end{cases}
\end{equation}
where $\beta$, the Stefan number  is given $\beta = L/(c\Delta T )$, where $c_s$ is the specific heat capacity of the solid phase. Also $\tilde{k}$ represents the ratio of thermal diffusivities of the solid and liquid phases respectively, $\tilde{k} = k_l/k_s$. Eq. \ref{theory3D} is solved using finite difference method as given in \cite{mccue2008classical} and \cite{voller1981accurate}. We
descretized the spatial derivatives using a central difference approximation and the 
temporal derivatives using forward time Euler approximation and solved the system 
explicitly to obtain the melting interface position in time.

In order to motivate application to realistic materials, we present the results for different latent heat capacities spanning different orders of magnitude. 
Though the results presented are non-dimensionalized, the 
material parameters are chosen to resemble realistic materials and these parameters will be used for further simulations in the dynamic context in the next section. In Fig. \ref{staticmelting3D} we present the time evolution of melting. Instead of the location of the interface position, we present the volume of the solid remaining at different instances in time. Since our goal is to 
present the effect of shape of the melting body, we choose the volume of the 
solid instead of a linear dimension. The simulation and discretization parameters are presented in Table \ref{table-interface3D}. Three different latent heat values are considered. 

We see that for lower latent heat values, the match is accurate, however for large latent heat values, the onset of melting is slightly different from the analytic solution. This is perhaps due to the truncated temperature and spatial kernel used in computing the latent heat. We will be addressing this 
in a future work. 

\begin{table}[h]
  \begin{center}
  \begin{tabular}{c c c}
\toprule
    Quantinty & Value & Unit \\
\midrule
    $\sigma $ & $0.2308$ & Nm\,$^{-1}$ \\ 
    $\Theta $ & $30$ & $^\circ$ \\
    $\eta $   & $0.001793$ & Pa\,s  \\
    $\rho $   & $1000$ & kg\,m$^{-3}$  \\
    $T_0 $    & $1$ & K  \\
    $T_m $    & $1.15$ & K  \\
    $T_\textrm{amb} $    & $4.0$ & K  \\
    $ C $    & $2110.0$ & J\,kg$^{-1}$\,K$^{-1}$  \\
    $ L $    & $0.1,1.0,100.0$ & kJ\,kg$^{-1}$  \\
    $ k $    & $2.14$ & W\,m$^{-1}$\,K$^{-1}$  \\
    $ \Delta x $    & $4.0\times10^{-5}$ & m  \\
    $ \Delta T $    & $0.1$ & K  \\
    $ R $    & $5.66\times 10^{-4}$ & m  \\
\bottomrule
  \end{tabular}
  \caption{Parameters used to simulate melting of a sphere and a milled particle.}
  \label{table-interface3D}
  \end{center}
\end{table}

The 
time update is numerically stable only when $\Delta t$ satisfies the condition 

\begin{equation}
  \Delta t \leq \min_a \left( 0.25\frac{h}{3|\vec{u}_a| } , 0.25\sqrt{\frac{m_a h}{3|\vec{f}^\text{int}_a|}}, 0.25\frac{\rho h^2}{9\mu} \right).
  \label{timestep}
\end{equation}

\section{Applications: Melting dynamics}
\label{results}

The multiphysics SPH algorithm is implemented in an ISPH code \cite{nair2014improved, nair2015volume} and melting problems with flow dynamics are solved for complex shaped three 
dimensional solids. Results are compared with theoretical results and other 
studies with simpler assumptions. 
\subsection{Melting of a mill particle}
\label{MeltingParticle}

\begin{figure}[!htb]
  \begin{center}
    \includegraphics[width=4in]{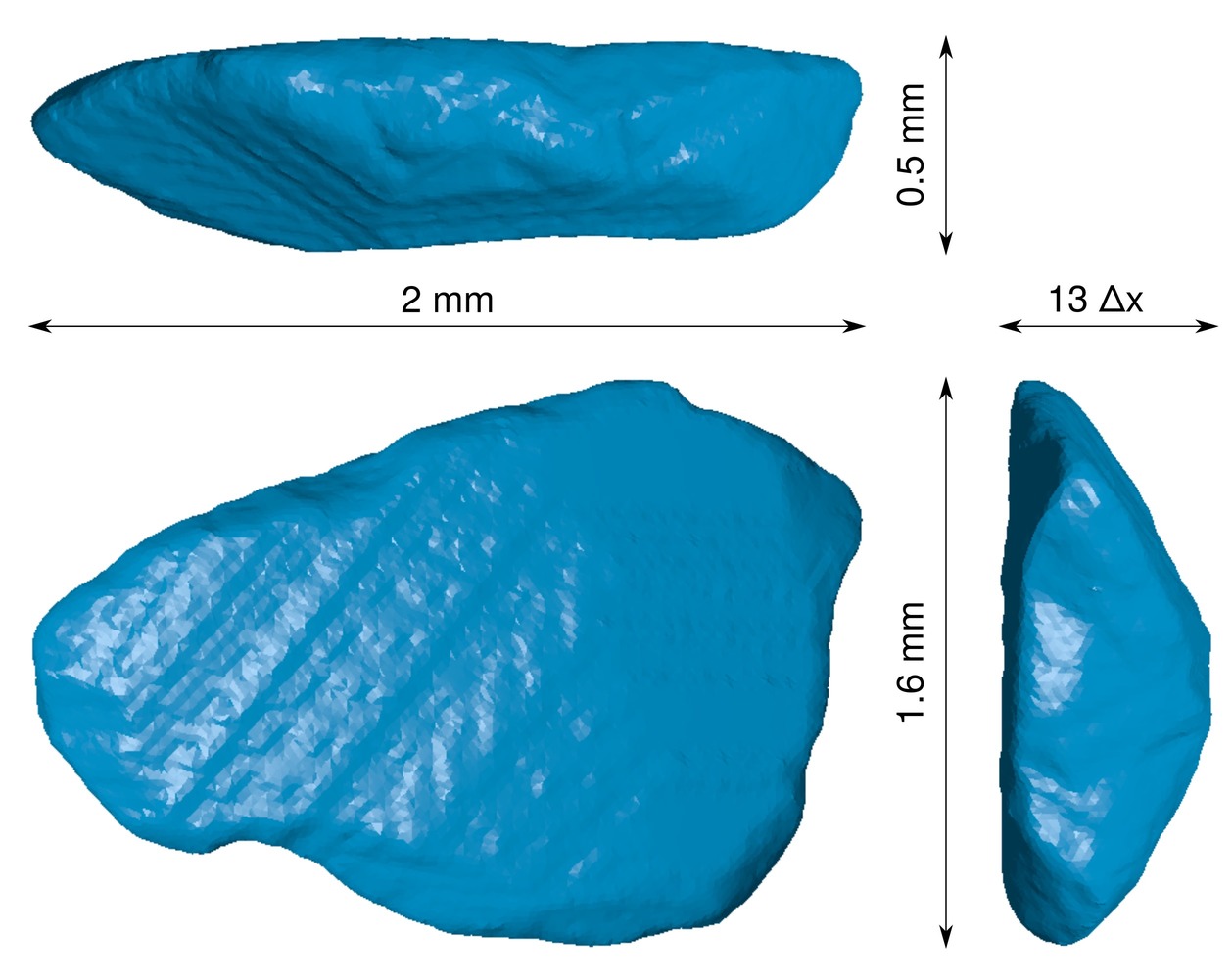}
    \caption{Complex shaped mill particle: dimensions in mm and in intial SPH particle spacing.}
    \label{milldimensions}
  \end{center}
\end{figure}

\begin{figure}[!htb]
  \begin{center}
\begin{subfigure}[t = 0]{1.5in}
  \includegraphics[width=1.5in,trim={1.2in 1.2in 1.2in 1.2in},clip]{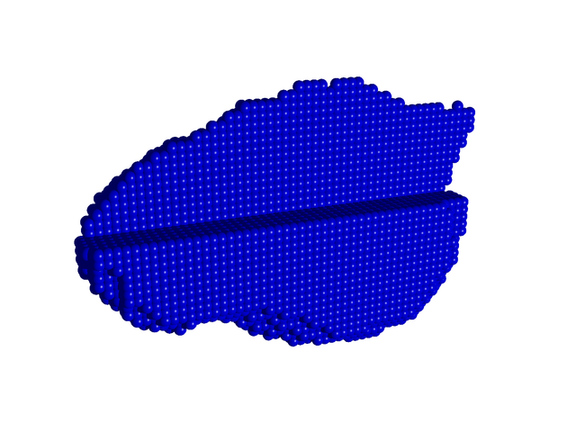}
\end{subfigure}
\begin{subfigure}[t=0.001\,s]{1.5in}
\includegraphics[width=1.5in,trim={1.2in 1.2in 1.2in 1.2in},clip]{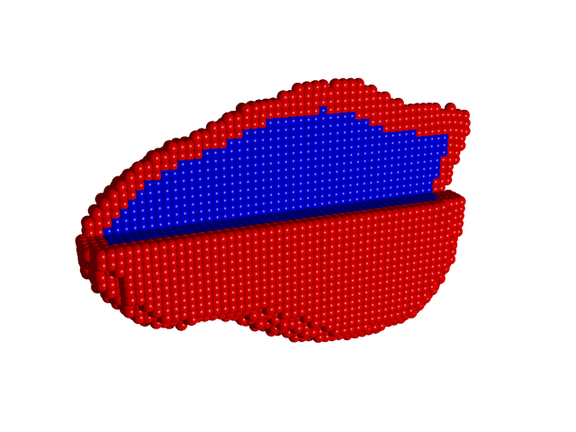}
\end{subfigure}
\begin{subfigure}[]{1.5in}
\includegraphics[width=1.5in,trim={1.2in 1.2in 1.2in 1.2in},clip]{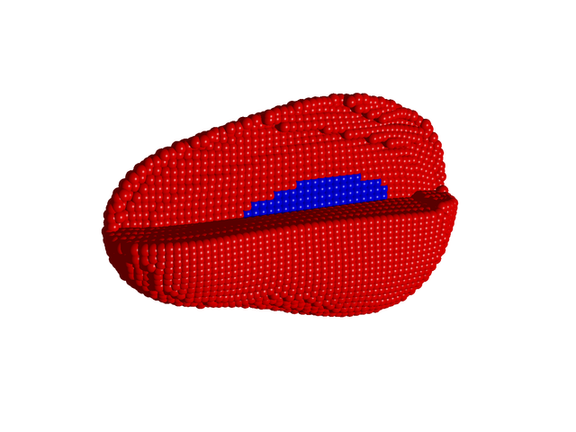}
\end{subfigure}
\begin{subfigure}[]{1.5in}
\includegraphics[width=1.5in,trim={1.2in 1.2in 1.2in 1.2in},clip]{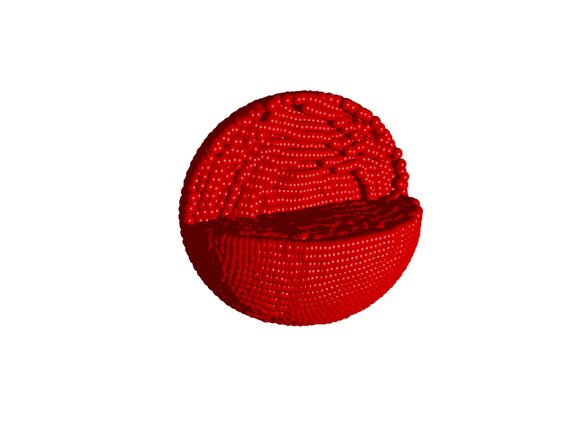}
\end{subfigure}
\begin{subfigure}[]{1.5in}
  \includegraphics[width=1.5in,trim={1.2in 1.2in 1.2in 1.2in},clip]{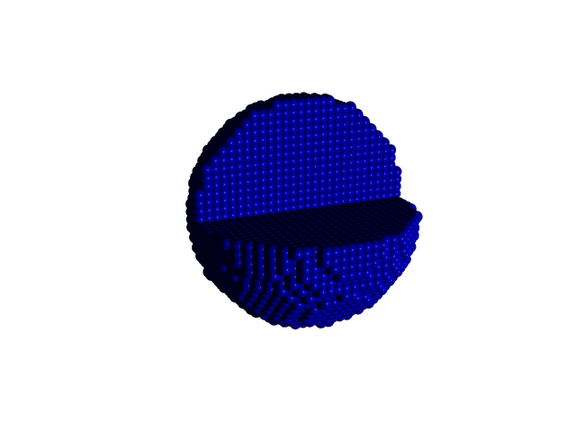}
			\caption{t=0\,s}
\end{subfigure}
\begin{subfigure}[]{1.5in}
  \includegraphics[width=1.5in,trim={1.2in 1.2in 1.2in 1.2in},clip]{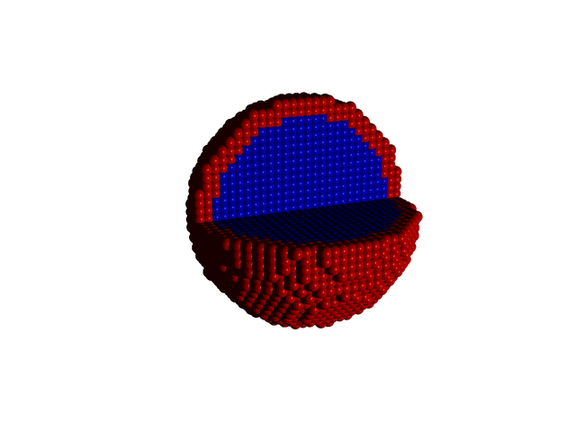}
			\caption{t=0.001\,s}
\end{subfigure}
\begin{subfigure}[]{1.5in}
  \includegraphics[width=1.5in,trim={1.2in 1.2in 1.2in 1.2in},clip]{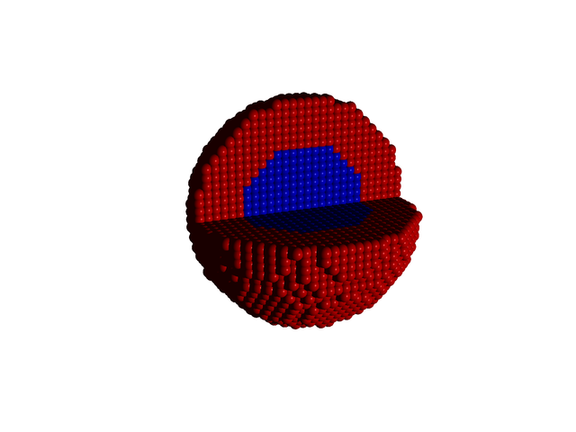}
			\caption{t=0.006\,s}
\end{subfigure}
\begin{subfigure}[]{1.5in}
  \includegraphics[width=1.5in,trim={1.2in 1.2in 1.2in 1.2in},clip]{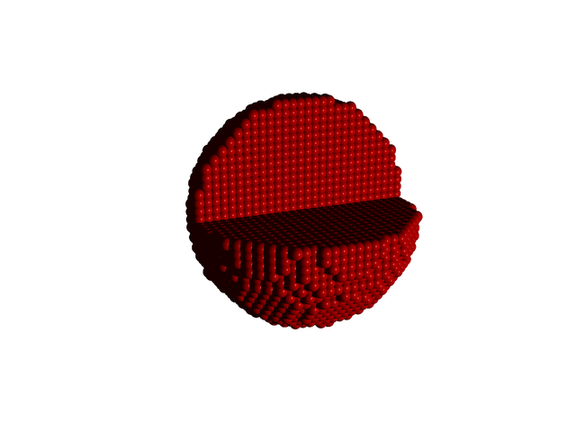}
			\caption{t=0.02\,s}
\end{subfigure}
  \caption{$L=0.01\,\textrm{kJ}/\textrm{kg}$: Melting of an irregular shaped mill particle vs a static sphere. Blue colored particles denote solid and red colored particles liquid state. }
	\label{MillPhase100}
  \end{center}
  \end{figure}\noindent

\begin{figure}[!htb]
  \begin{center}
\begin{subfigure}[t = 0]{1.5in}
\includegraphics[width=1.5in,trim={1.2in 1.2in 1.2in 1.2in},clip]{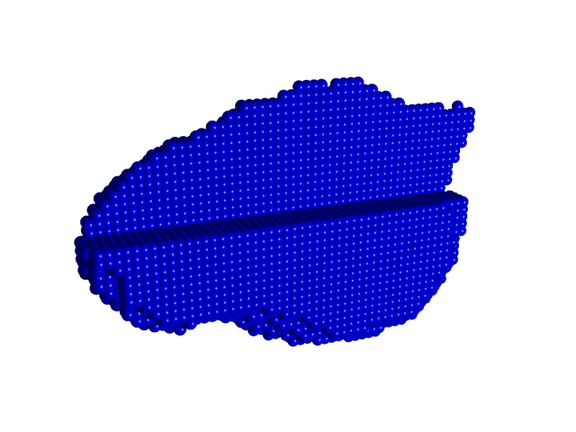}
\end{subfigure}
\begin{subfigure}[t=0.001\,s]{1.5in}
\includegraphics[width=1.5in,trim={1.2in 1.2in 1.2in 1.2in},clip]{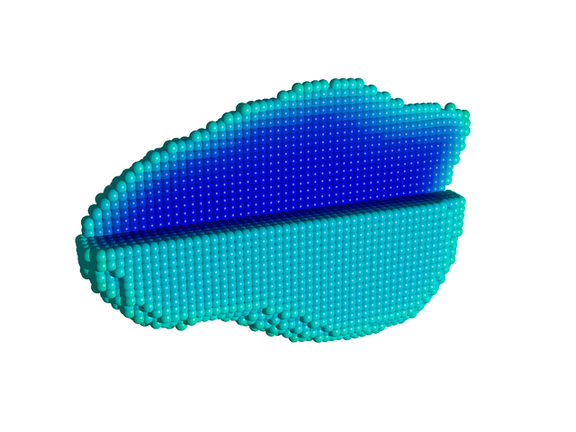}
\end{subfigure}
\begin{subfigure}[]{1.5in}
\includegraphics[width=1.5in,trim={1.2in 1.2in 1.2in 1.2in},clip]{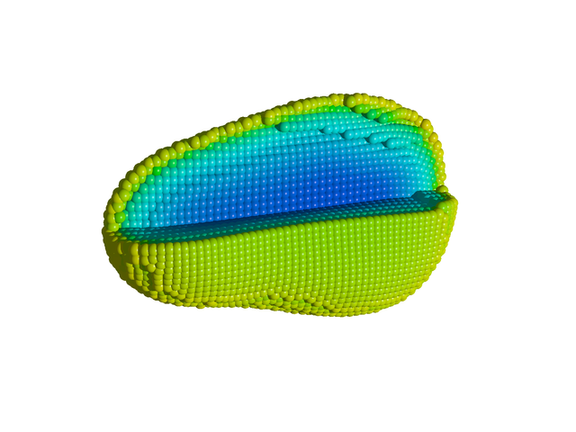}
\end{subfigure}
\begin{subfigure}[]{1.5in}
\includegraphics[width=1.5in,trim={1.2in 1.2in 1.2in 1.2in},clip]{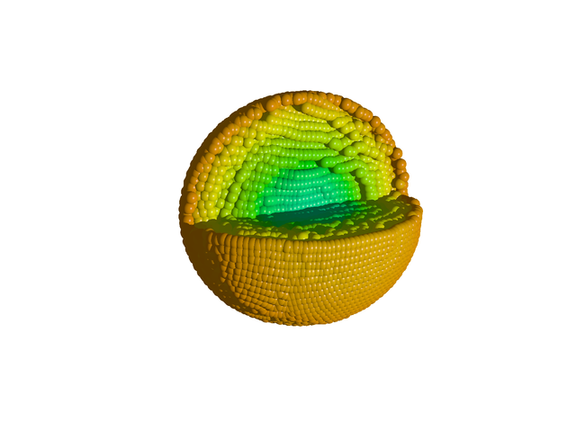}
\end{subfigure}
\begin{subfigure}[]{1.5in}
  \includegraphics[width=1.5in,trim={1.2in 1.2in 1.2in 1.2in},clip]{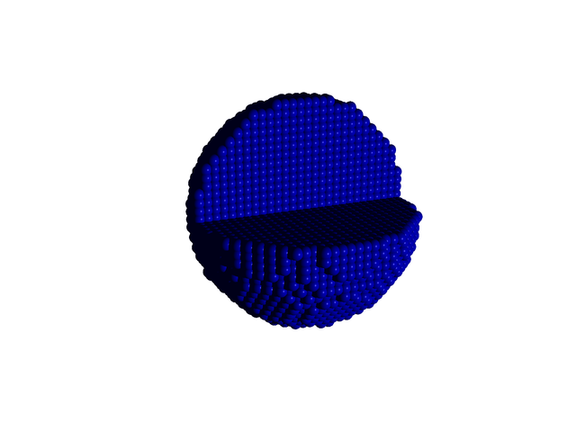}
			\caption{t=0\,s}
\end{subfigure}
\begin{subfigure}[]{1.5in}
  \includegraphics[width=1.5in,trim={1.2in 1.2in 1.2in 1.2in},clip]{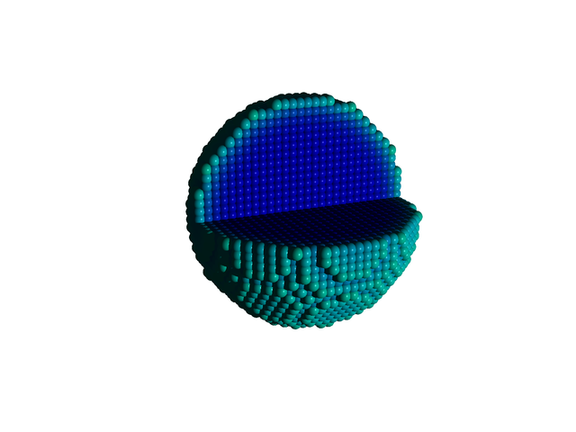}
			\caption{t=0.001\,s}
\end{subfigure}
\begin{subfigure}[]{1.5in}
  \includegraphics[width=1.5in,trim={1.2in 1.2in 1.2in 1.2in},clip]{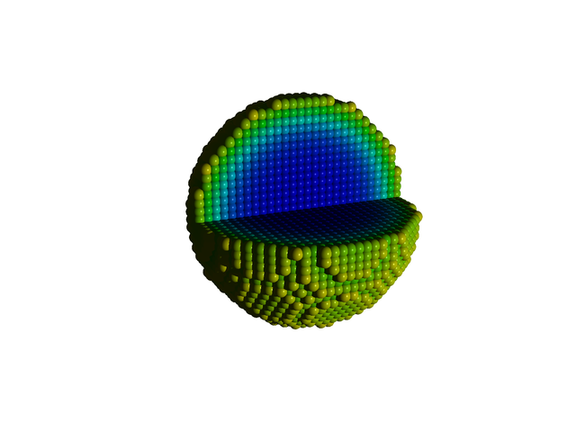}
			\caption{t=0.006\,s}
\end{subfigure}
\begin{subfigure}[]{1.5in}
  \includegraphics[width=1.5in,trim={1.2in 1.2in 1.2in 1.2in},clip]{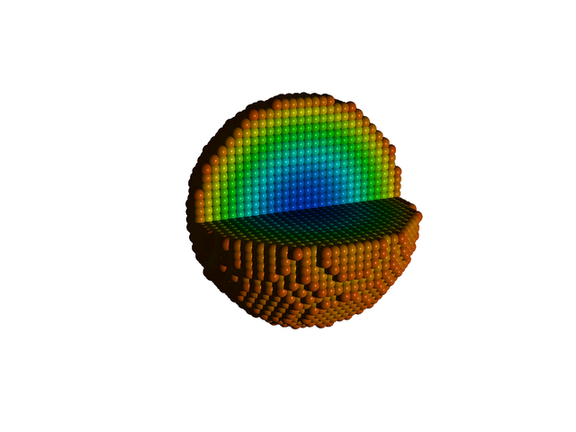}
			\caption{t=0.02\,s}
\end{subfigure}
  \caption{$L=0.01\,\textrm{kJ}/\textrm{kg}$: Temperature distribution in 
  the melting mill particle.}

\label{MillTemp100}

  \end{center}
  \end{figure}

\begin{figure}[!htb]
  \begin{center}
\begin{subfigure}[t = 0]{1.5in}
\includegraphics[width=1.5in,trim={1.2in 1.2in 1.2in 1.2in},clip]{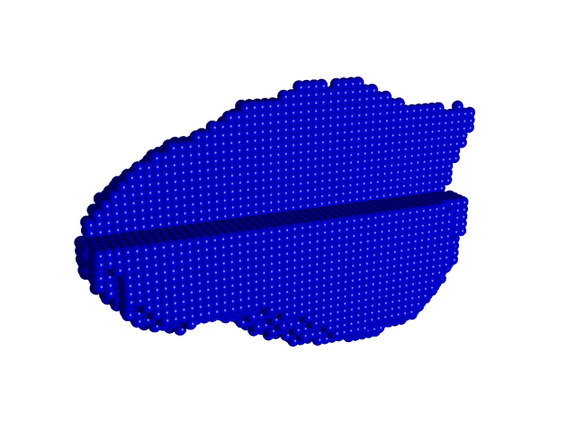}
\end{subfigure}
\begin{subfigure}[t=0.001\,s]{1.5in}
\includegraphics[width=1.5in,trim={1.2in 1.2in 1.2in 1.2in},clip]{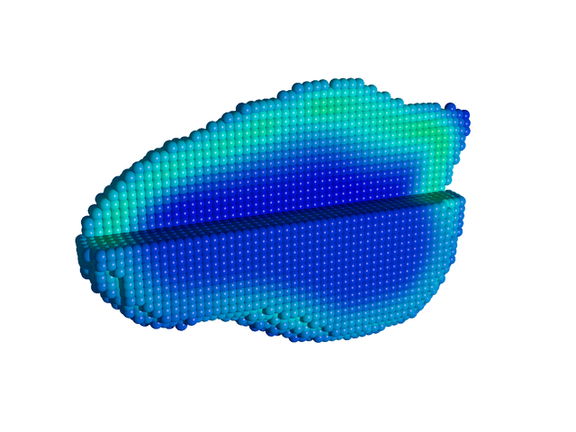}
\end{subfigure}
\begin{subfigure}[]{1.5in}
\includegraphics[width=1.5in,trim={1.2in 1.2in 1.2in 1.2in},clip]{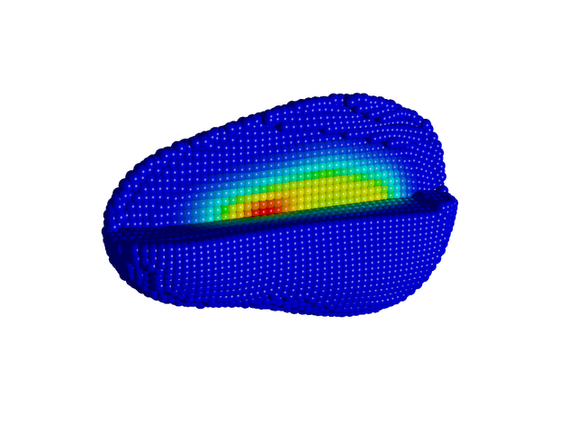}
\end{subfigure}
\begin{subfigure}[]{1.5in}
\includegraphics[width=1.5in,trim={1.2in 1.2in 1.2in 1.2in},clip]{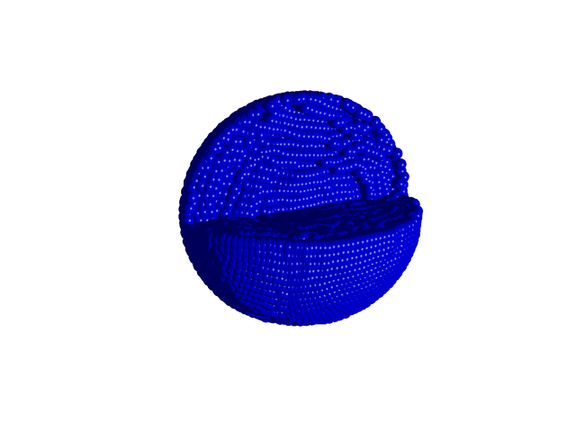}
\end{subfigure}
\begin{subfigure}[]{1.5in}
  \includegraphics[width=1.5in,trim={1.2in 1.2in 1.2in 1.2in},clip]{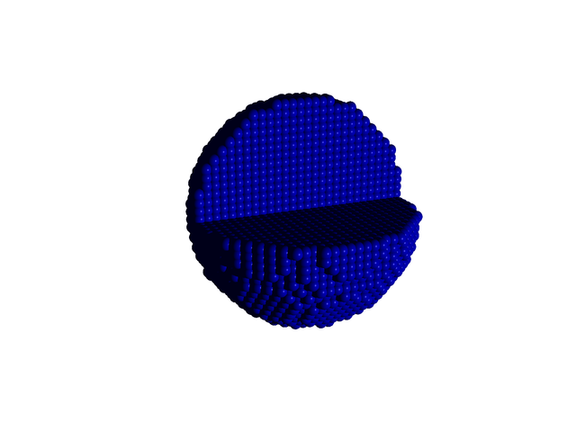}
			\caption{t=0\,s}
\end{subfigure}
\begin{subfigure}[]{1.5in}
  \includegraphics[width=1.5in,trim={1.2in 1.2in 1.2in 1.2in},clip]{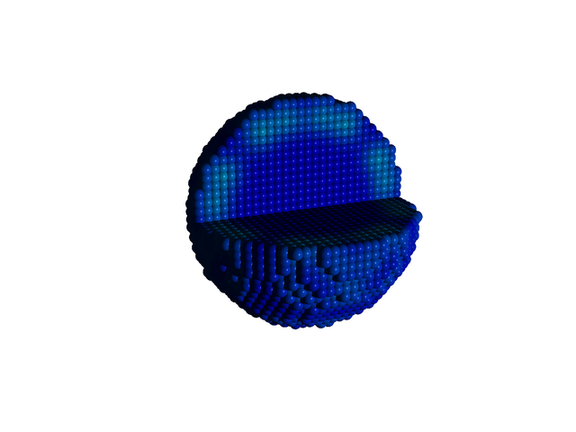}
			\caption{t=0.001\,s}
\end{subfigure}
\begin{subfigure}[]{1.5in}
  \includegraphics[width=1.5in,trim={1.2in 1.2in 1.2in 1.2in},clip]{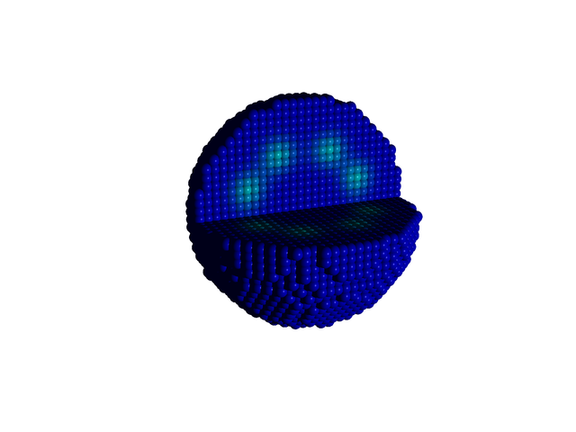}
			\caption{t=0.006\,s}
\end{subfigure}
\begin{subfigure}[]{1.5in}
  \includegraphics[width=1.5in,trim={1.2in 1.2in 1.2in 1.2in},clip]{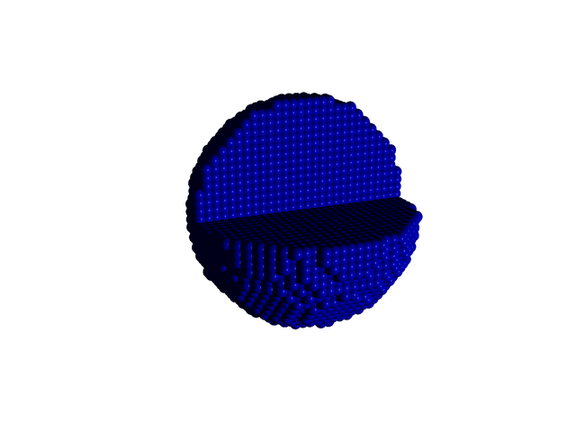}
			\caption{t=0.02\,s}
\end{subfigure}
  \caption{$L=0.01\,\textrm{kJ}/\textrm{kg}$: Latent heat absorbed at the 
  interface during the melting of the mill particle.}

  \label{MillLat100}
  \end{center}
  \end{figure}


\begin{figure}[!htb]
  \begin{center}
\begin{subfigure}[t = 0]{1.5in}
\includegraphics[width=1.5in,trim={1.2in 1.2in 1.2in 1.2in},clip]{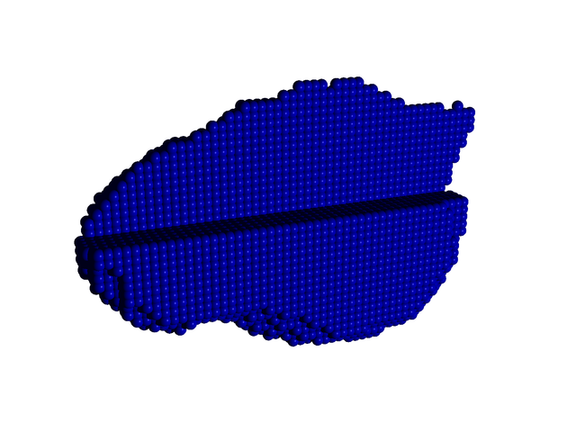}
\end{subfigure}
\begin{subfigure}[t=0.001\,s]{1.5in}
\includegraphics[width=1.5in,trim={1.2in 1.2in 1.2in 1.2in},clip]{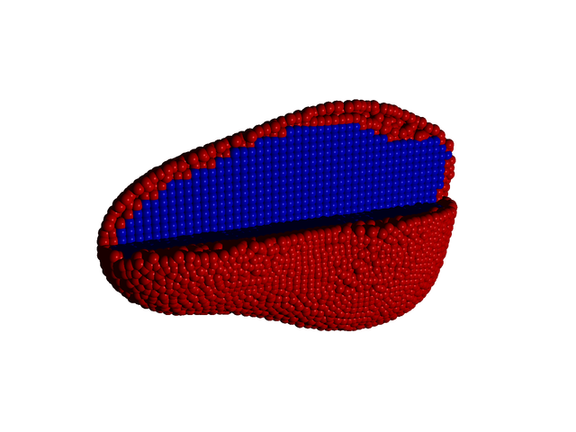}
\end{subfigure}
\begin{subfigure}[]{1.5in}
\includegraphics[width=1.5in,trim={1.2in 1.2in 1.2in 1.2in},clip]{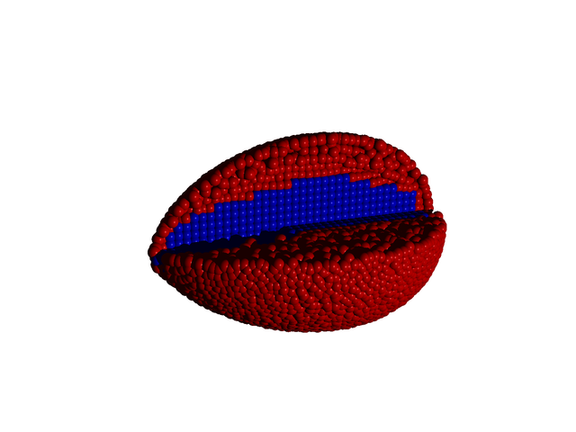}
\end{subfigure}
\begin{subfigure}[]{1.5in}
\includegraphics[width=1.5in,trim={1.2in 0.8in 1.2in 1.6in},clip]{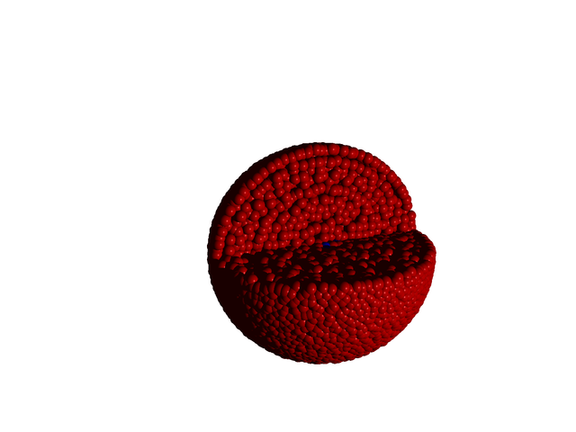}
\end{subfigure}
\begin{subfigure}[]{1.5in}
  \includegraphics[width=1.5in,trim={1.2in 1.2in 1.2in 1.2in},clip]{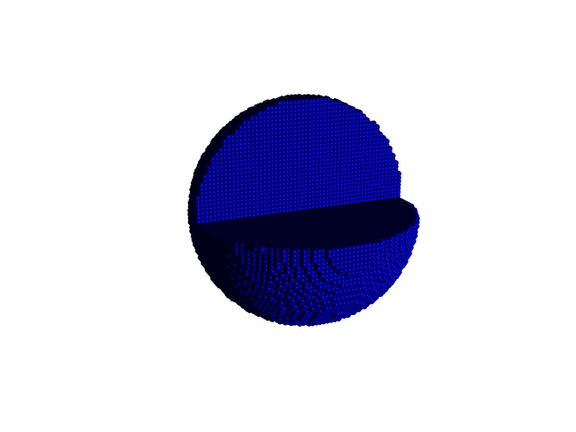}
			\caption{t=0\,s}
\end{subfigure}
\begin{subfigure}[]{1.5in}
  \includegraphics[width=1.5in,trim={1.2in 1.2in 1.2in 1.2in},clip]{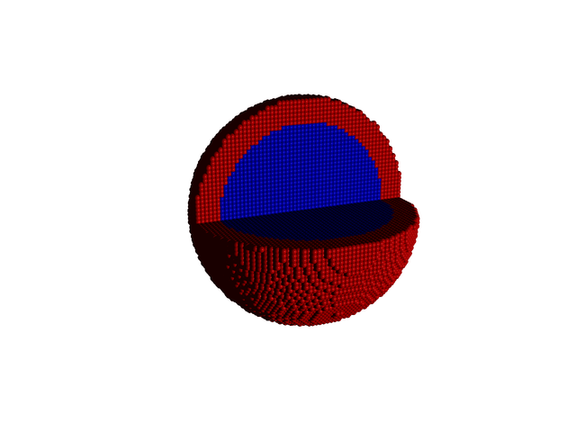}
			\caption{t=0.10\,s}
\end{subfigure}
\begin{subfigure}[]{1.5in}
  \includegraphics[width=1.5in,trim={1.2in 1.2in 1.2in 1.2in},clip]{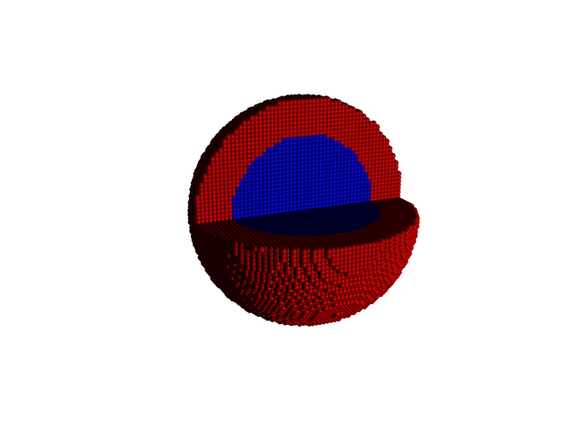}
			\caption{t=0.19\,s}
\end{subfigure}
\begin{subfigure}[]{1.5in}
  \includegraphics[width=1.5in,trim={1.2in 1.2in 1.2in 1.2in},clip]{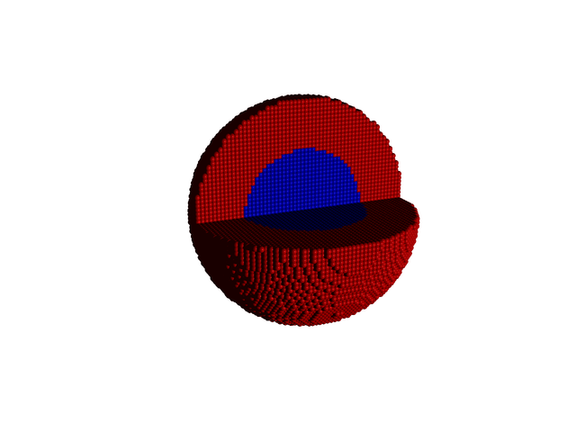}
			\caption{t=0.32\,s}
\end{subfigure}
  \caption{$L=100.0\,\textrm{kJ}/\textrm{kg}$: Melting of an irregular shaped mill particle vs a static sphere. Blue colored particles denote solid and red colored particles liquid state.}

  \label{MillPhase100000}
  \end{center}
  \end{figure}

\begin{figure}[!htb]
  \begin{center}
\begin{subfigure}[t = 0]{1.5in}
\includegraphics[width=1.5in,trim={1.2in 1.2in 1.2in 1.2in},clip]{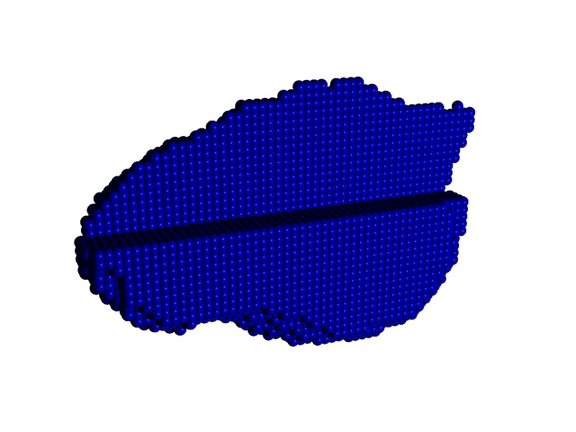}
\end{subfigure}
\begin{subfigure}[t=0.001\,s]{1.5in}
\includegraphics[width=1.5in,trim={1.2in 1.2in 1.2in 1.2in},clip]{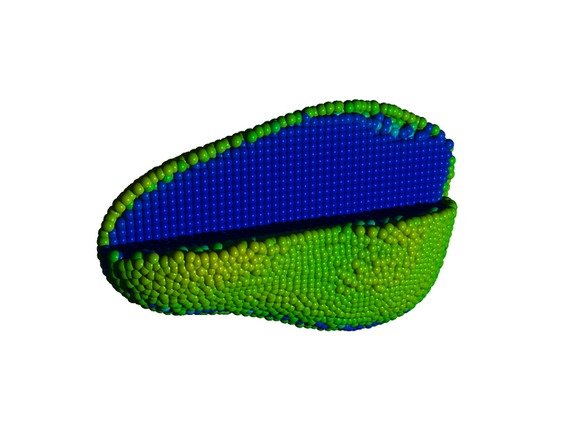}
\end{subfigure}
\begin{subfigure}[]{1.5in}
\includegraphics[width=1.5in,trim={1.2in 1.2in 1.2in 1.2in},clip]{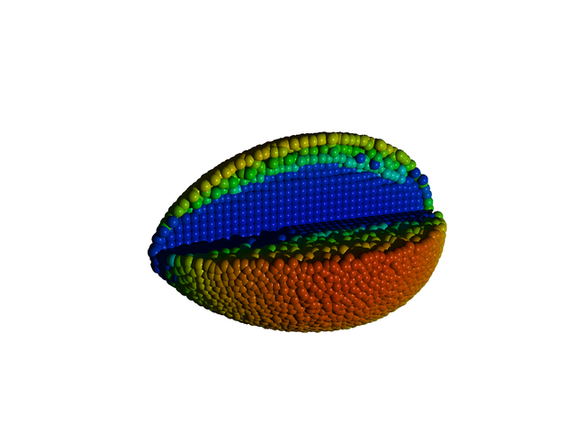}
\end{subfigure}
\begin{subfigure}[]{1.5in}
\includegraphics[width=1.5in,trim={1.2in 0.8in 1.2in 1.6in},clip]{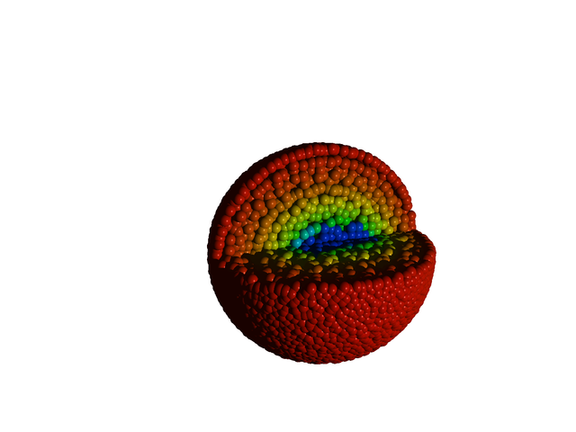}
\end{subfigure}
\begin{subfigure}[]{1.5in}
  \includegraphics[width=1.5in,trim={1.2in 1.2in 1.2in 1.2in},clip]{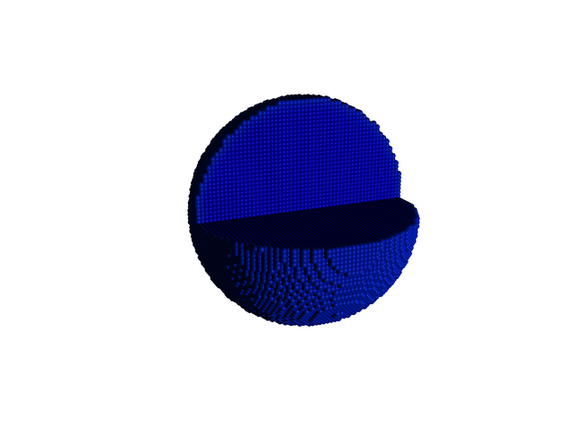}
			\caption{t=0\,s}
\end{subfigure}
\begin{subfigure}[]{1.5in}
  \includegraphics[width=1.5in,trim={1.2in 1.2in 1.2in 1.2in},clip]{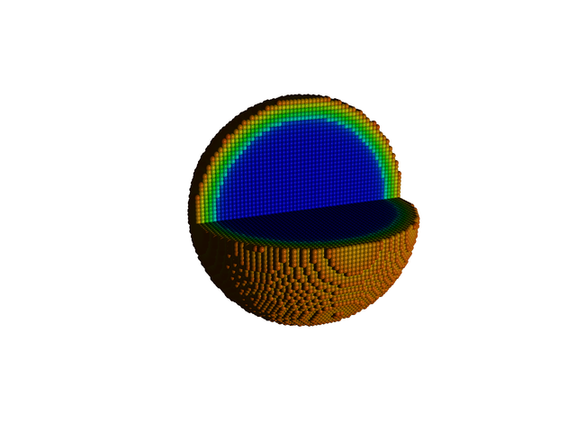}
			\caption{t=0.10\,s}
\end{subfigure}
\begin{subfigure}[]{1.5in}
  \includegraphics[width=1.5in,trim={1.2in 1.2in 1.2in 1.2in},clip]{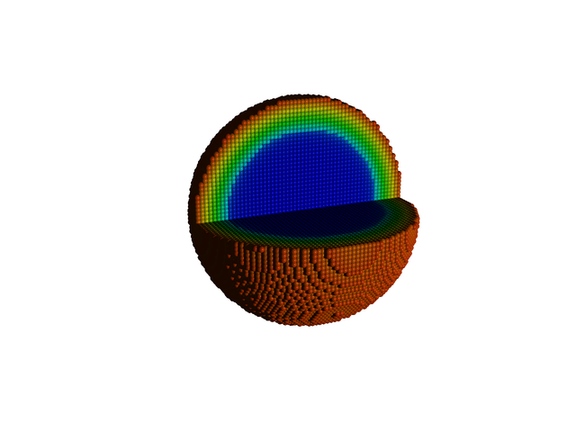}
			\caption{t=0.19\,s}
\end{subfigure}
\begin{subfigure}[]{1.5in}
  \includegraphics[width=1.5in,trim={1.2in 1.2in 1.2in 1.2in},clip]{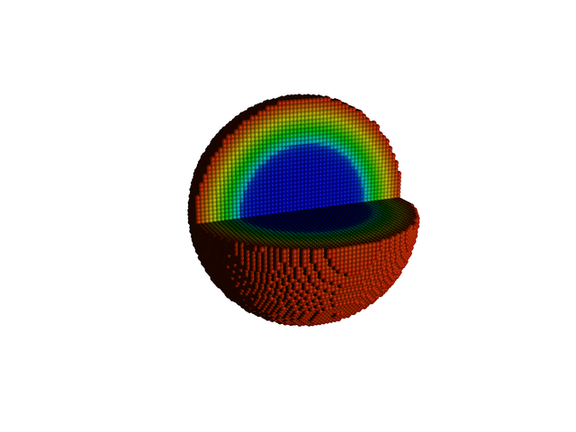}
			\caption{t=0.32\,s}
\end{subfigure}
  \caption{$L=100.0\,\textrm{kJ}/\textrm{kg}$:Temperature field in 
  the melting particles. Top row shows the mill particle with flow of melt and the bottom row shows 
  a static spherical particle.}

  \label{MillTemp100000}
  \end{center}
  \end{figure}
\begin{figure}[!htb]
  \begin{center}
\begin{subfigure}[t = 0]{1.5in}
\includegraphics[width=1.5in,trim={1.2in 1.2in 1.2in 1.2in},clip]{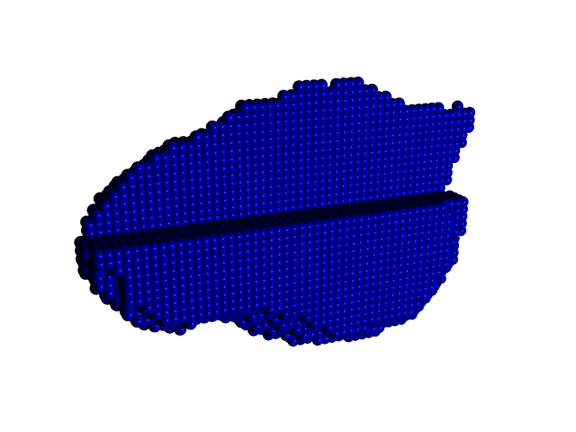}
\end{subfigure}
\begin{subfigure}[t=0.10\,s]{1.5in}
\includegraphics[width=1.5in,trim={1.2in 1.2in 1.2in 1.2in},clip]{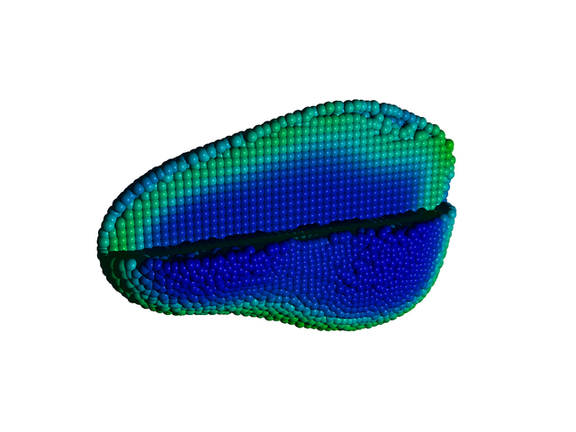}
\end{subfigure}
\begin{subfigure}[]{1.5in}
\includegraphics[width=1.5in,trim={1.2in 1.2in 1.2in 1.2in},clip]{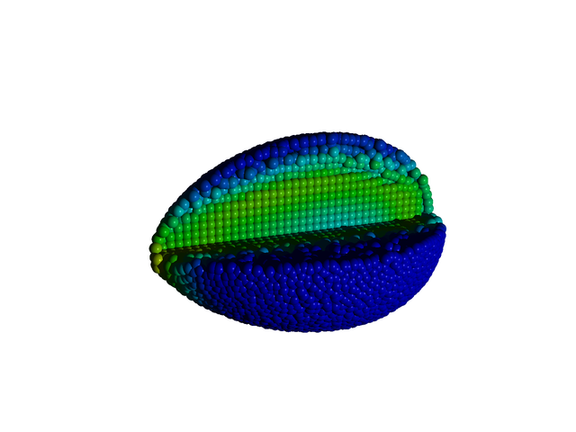}
\end{subfigure}
\begin{subfigure}[]{1.5in}
\includegraphics[width=1.5in,trim={1.2in 0.8in 1.2in 1.6in},clip]{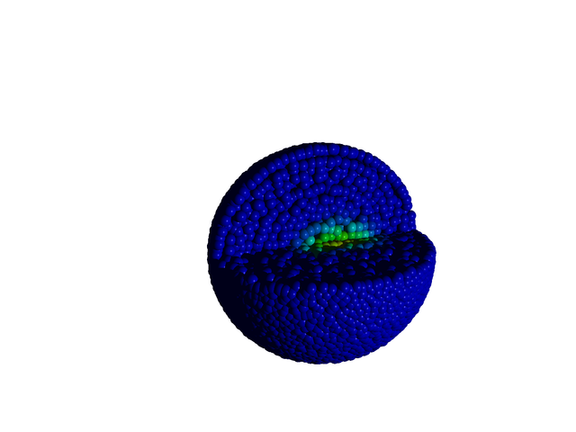}
\end{subfigure}
\begin{subfigure}[]{1.5in}
  \includegraphics[width=1.5in,trim={1.2in 1.2in 1.2in 1.2in},clip]{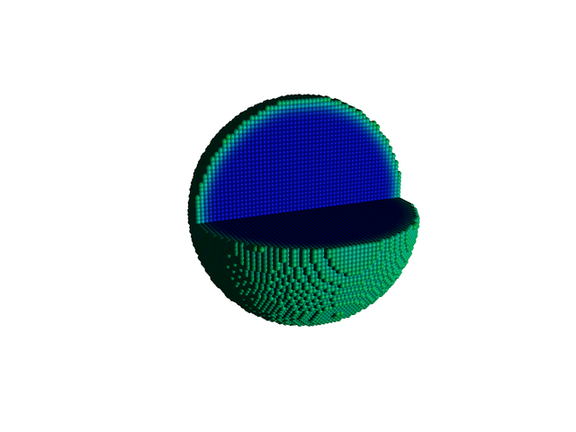}
			\caption{t=0\,s}
\end{subfigure}
\begin{subfigure}[]{1.5in}
  \includegraphics[width=1.5in,trim={1.2in 1.2in 1.2in 1.2in},clip]{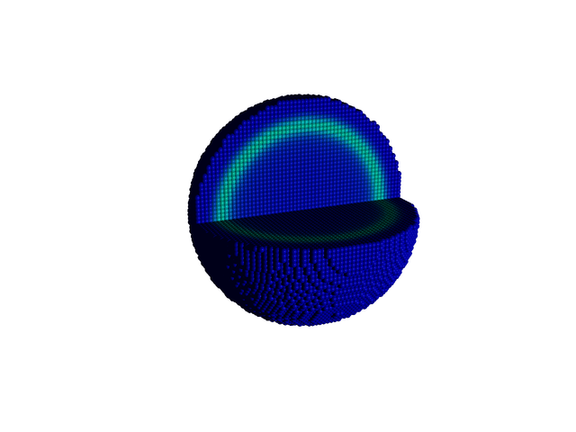}
			\caption{t=0.10\,s}
\end{subfigure}
\begin{subfigure}[]{1.5in}
  \includegraphics[width=1.5in,trim={1.2in 1.2in 1.2in 1.2in},clip]{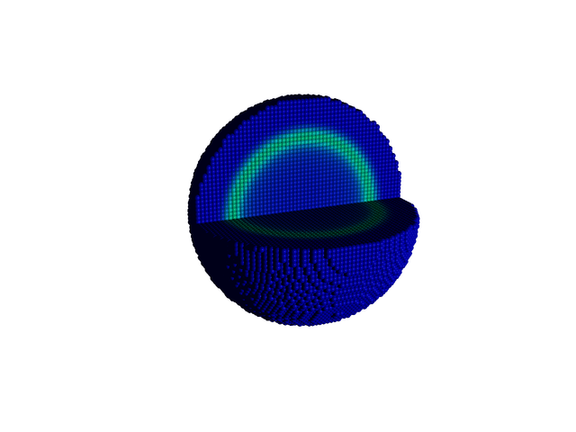}
			\caption{t=0.19\,s}
\end{subfigure}
\begin{subfigure}[]{1.5in}
  \includegraphics[width=1.5in,trim={1.2in 1.2in 1.2in 1.2in},clip]{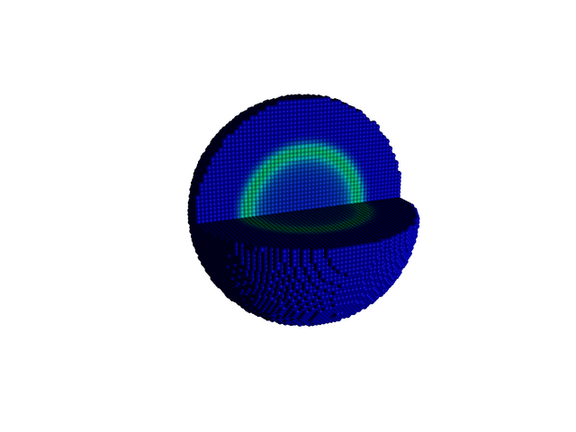}
			\caption{t=0.32\,s}
\end{subfigure}
  \caption{$L=100.0\,\textrm{kJ}/\textrm{kg}$:Latent heat absorbed at the 
  interface during the melting of the mill particle.}

  \label{MillLat100000}
  \end{center}
  \end{figure}

\begin{figure}[!htb]
  \begin{center}
    \includegraphics[width=5in]{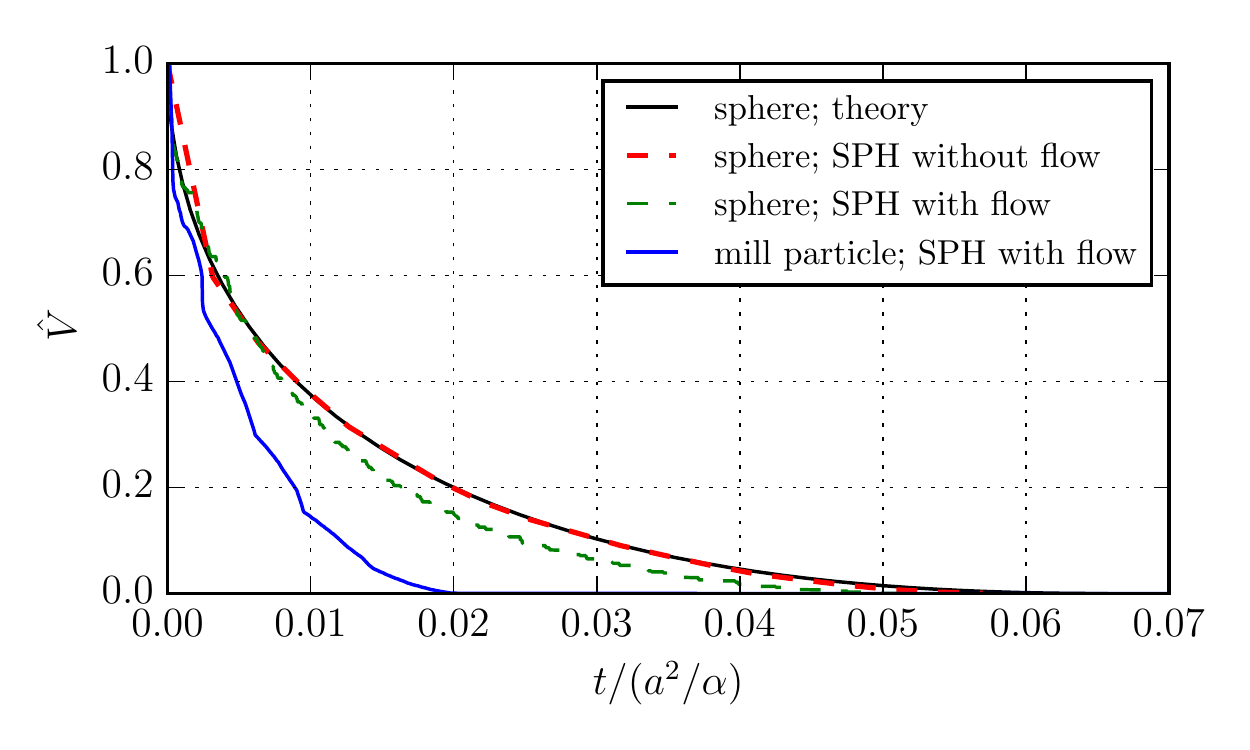}
    \caption{Time variation of volume fraction of solid for a sphere and the mill particle of equal volume.}
    \label{volfrac100}
  \end{center}
\end{figure}
\begin{figure}[!htb]
  \begin{center}
    \includegraphics[width=5in]{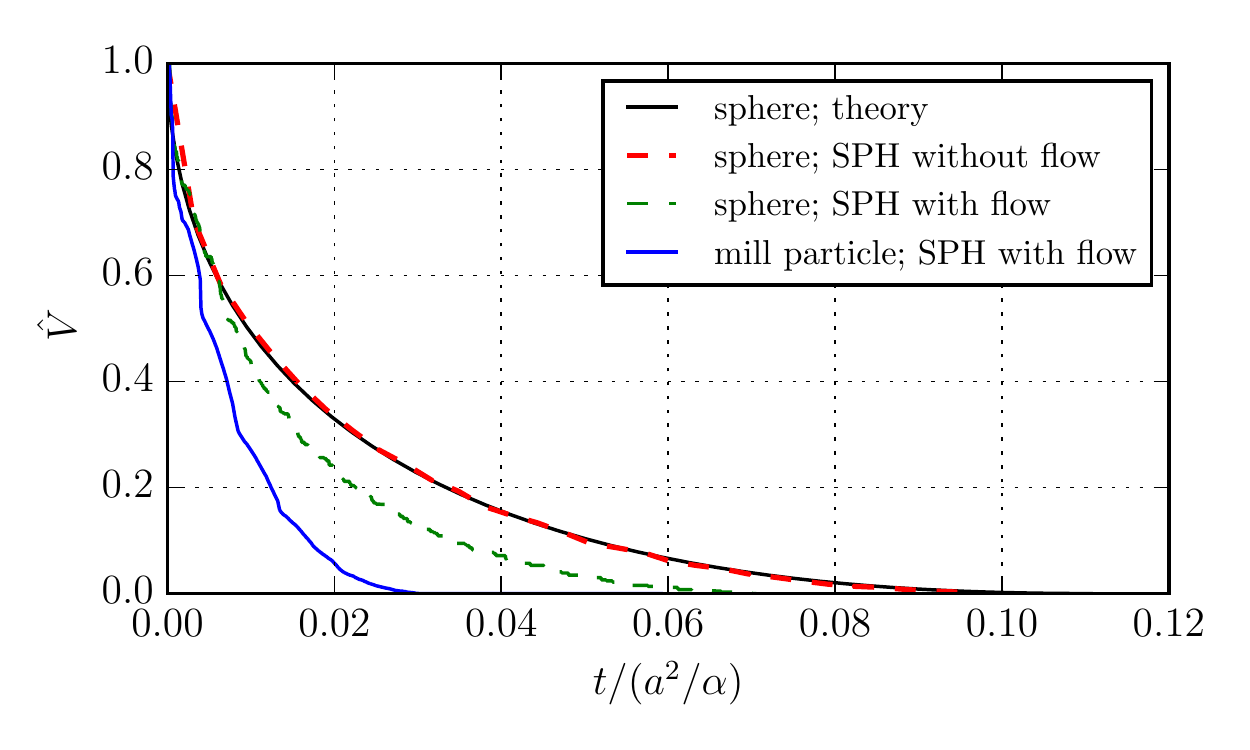}
    \caption{Time variation of volume fraction of solid for a sphere and the mill particle of equal volume.}
    \label{volfrac1000}
  \end{center}
\end{figure}
\begin{figure}[!htb]
  \begin{center}
    \includegraphics[width=5in]{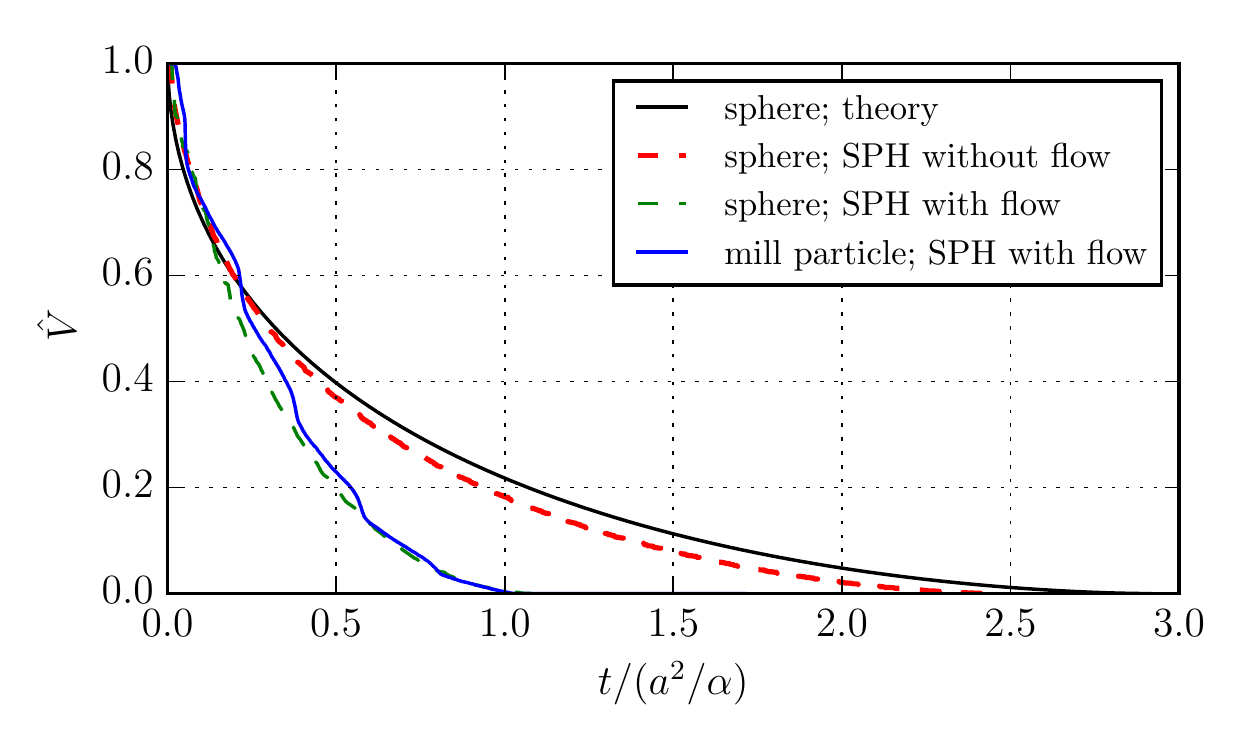}
    \caption{Time variation of volume fraction of solid for a sphere and the mill particle of equal volume.}
    \label{volfrac100000}
  \end{center}
\end{figure}

A complex shaped geometry obtained by scanning a milled particle from an 
industrial process is shown in Fig. \ref{milldimensions}. This geometry is 
filled with SPH particles in a square array with the same intial spacing and 
simulation parameters as that is used in 
the validations using melting of a sphere presented in the previous section. 
Even the volume of the sphere in the previous section was chosen to be equal to 
the volume of the mill particle. We have avoided discontinuities in the thermal 
properties across the solid liquid interface in order to compare with the 
theoretical results and to focus on the melting and capillary force based 
deformation as the first step. Discontinuities in this regard will be introduced
in the next section. 

The evolution of the solid and liquid phases are shown in Fig. \ref{MillPhase100}--\ref{MillLat100000}. Since different 
physical processes are coupled in this simulation, we have not attempted to non-dimensionalize
this system. However the rate of phase change is non dimensionalized and compared 
with the melting of the sphere in the earlier section.

In Figs. \ref{MillPhase100} and \ref{MillPhase100000} blue and red particles represent solid and liquid phases, respectively. For better illustration, a quadrant is clipped out of the irregular particle to show the interface evolution in the bulk of the particle. At the beginning of the simulation, all {SPH} particles are in solid state. As a result of the heat exchange with the ambiance, the phase change occurs as a particle temperature rises above the melting point. A layer with liquid {SPH} particles are formed on top of the surface of the irregular shaped particle after about $0.001\,s$. 
The liquid solid interface is hydrophilic owing to the acute contact angle of $ 30^{\circ} $, resulting in a wetted surface. The flow of the liquid due to capillary forces influences the heat exchange due to the evolving geometry.

Shortly after a thin layer is melted, the shape of the body smoothes out by $0.005\,s$. We have presented the simulation until all the solid is finally melted. After the solid is completely 
melted, the resulting liquid drop performs osciallations until relaxation to a spherical
shape of minimum surface area.

In Figs. \ref{MillTemp100} and \ref{MillTemp100000}, the temperature field of 
the melting body is shown. 
With the onset of heat exchange with the ambiance the initial solid particles start to heat up. At 0.001\,s the particles have the same temperature depending on the exposure to free surface. Due to the convective movement of the particles caused by surface tension, liquid particles are moving to the vicinity of the solid body, where they cool down. It is also possible that colder liquid particles are transported back to the surface where they are warmed up.
This results in an irregular distribution of particles with different temperatures at a certain time which can already be seen after 0.013\,s and is further strengthened during the melting process until all particles are completely in liquid state.

In Figs. \ref{MillLat100} and \ref{MillLat100000}, the area of effect of the normalized latent heat is shown.
As the temperature rises, particles enter the mushy region where the effect of latent heat increases. Red colored particles characterize a high effect of latent heat and blue colored particles characterize particles that are not influenced by the latent heat. It can be seen that the edges of the irregular shaped particle are first influenced by the phase change. Since the still unfused rigid body moves to the upper edge of the particle, the effect of the latent heat remains exclusively on the upper hemisphere of the sphere formed around the solid body.

The same simulation with properties given in Table \ref{table-interface3D} was performed for a latent heat of $L=100\,\nicefrac{\mathrm{kJ}}{\mathrm{kg}}$ 
The process of phase change is shown in Fig. \ref{MillPhase100}. In this case, the phase change happens slower than the movement of liquid particles due to surface tension. It takes 0.40\,s until all solid particles are melted. In comparison to a latent heat of $L=1\,\nicefrac{\mathrm{kJ}}{\mathrm{kg}}$ with a melting time 0.037\,s this is approximately ten times slower. It can be seen that the liquid particles are moving behind the rigid body  at $t=0.15\,\mathrm{s}$ where they are forming a sphere. One reason for this could be an unequal melting rate on both sides of the solid body. Therefore, the liquid particles are moving to the side with more liquid particles at a specific time. Another reason could be the infiltration of the liquid particles through the rigid body, which is a numerical issue. 
In comparison to the previous case with $L=1\,\nicefrac{\mathrm{kJ}}{\mathrm{kg}}$ the rotation of the rigid body within the forming liquid sphere is increased as shown in the time interval 0.15\,s to 0.25\,s. This effect results from the fact that the solid particles are melting more slowly. That is why the exchange of forces between liquid particles and rigid body is not altered by the phase change process as much as for a lower latent heat number. As the rigid body shrinks it is moving towards the liquid spheres center which is in contrast to the case for the latent heat of $L=1\, \nicefrac{\mathrm{kJ}}{\mathrm{kg}}$ as shown in Fig. \ref{MillPhase100000}. 

In Fig. \ref{MillTemp100000} the temperature distribution for each investigated time is shown. At the beginning, the irregular shaped particle has an uniform initial temperature of $1.00\,\mathrm{K}$. As the time proceeds, the temperature rises at the solid bodies surface.  Due to the latent heat there is a high temperature gradient between solid and liquid particles as it can be seen at $t=0.15\,\mathrm{s}$. After $t=0.40\,\mathrm{s}$ the maximum temperature occupied by a liquid particle is 3.43\,K and hence approximately $1\,K$ above the highest temperature in the previous case with $L=1.00\,\mathrm{kJ}$. 
Because the time needed for melting increases with the latent heat, the liquid particles temperature tends to reach an equilibrium value as shown in the time interval between 0.3\,s and 0.4\,s. 

The normalized latent heat for the irregular shaped particle with $L=1\,\nicefrac{\mathrm{kJ}}{\mathrm{kg}}$ can be seen in Fig. \ref{MillLat100000}. One difference to Fig. \ref{MillLat100} is the propagation of the areas where latent heat comes into action which is caused by the rigid bodies position.  

The evolution of solid fraction with time for all three latent heat values are show in Figs. \ref{volfrac100}, \ref{volfrac1000}, \ref{volfrac100000}. Here the analytical and static sphere results are the same as in Fig. \ref{staticmelting3D}. We have also included dynamic melting of spheres in these figures in order to detect increased heat transfer due flows due to instabilities. The increased heat transfer of a dynamic sphere suggests the presents of capillary flows, which exacerbates for larger latent heat values.
  
\subsection{Agglomeration of a chain of melting solids} 
\begin{figure}[!htb]
	\begin{center}	
    {\renewcommand{\arraystretch}{1.25}
    \begin{tabular}{c>{\centering\arraybackslash}p{0.5in}cc>{\centering\arraybackslash}p{1in}cc}
      $\hat{t} $ & VS & explicit & implicit & VS & explicit & impliciti \\

      {$0$}  & \multicolumn{3}{c}{\multirow{7}{*}{ \includegraphics[trim=0.4in 0 0 0, clip, height=1.8in]{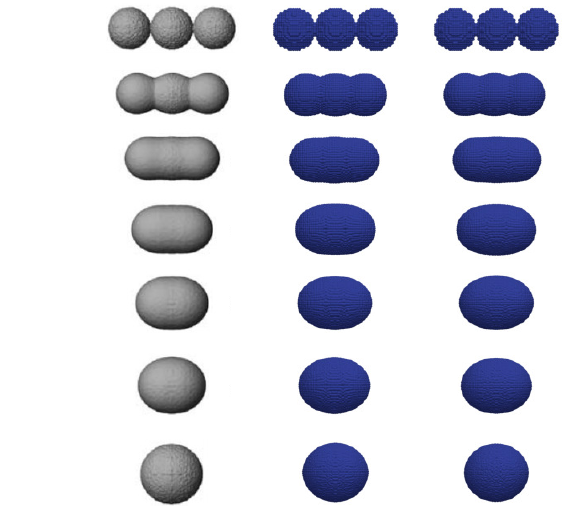}}} & \multicolumn{3}{c}{\multirow{7}{*}{\includegraphics[height=1.8in]{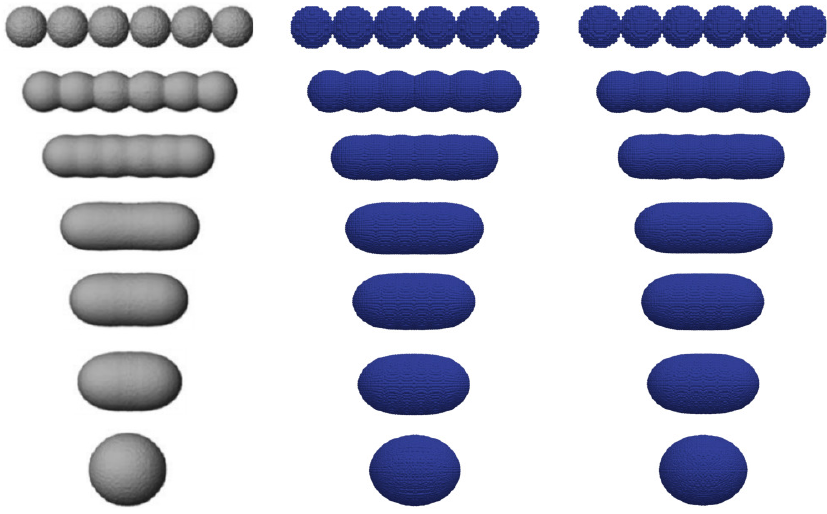}}} \\
      {$0.91$} &\multicolumn{3}{c}{}&\multicolumn{3}{c}{}\\  
      {$1.82$} &\multicolumn{3}{c}{}&\multicolumn{3}{c}{}\\    
      {$2.72$} &\multicolumn{3}{c}{}&\multicolumn{3}{c}{}\\    
      {$3.62$} &\multicolumn{3}{c}{}&\multicolumn{3}{c}{}\\    
      {$4.52$} &\multicolumn{3}{c}{}&\multicolumn{3}{c}{}\\   
      {$9.05$} &\multicolumn{3}{c}{}&\multicolumn{3}{c}{}
    \end{tabular} }
	\caption{Shape evolution of agglomerate chains with three and six spheres. On the left side of each subfigure are the solutions by \cite{kirchhof2009three}. The centered column are the solutions with an explicit  and the right side with an implicit method.}
	\label{KirchhoffComp}
  \end{center}
\end{figure}

\begin{figure}[!htb]
  \begin{center}
	\begin{tikzpicture}
	\begin{axis}[ 
	width=4in,
	ymin=0.0,
	xlabel = $t \sigma \left(\eta S_0\right)^{-1}$,
	ylabel = $S S_0^{-1}$,
	xtick pos=left,
	ytick pos=left,
	legend cell align={left},
	ymin =0.5, ymax = 1.0,
	restrict y to domain=-0.01:1.1, restrict x to domain=0:9.1,
	legend pos = north east,
	legend style={draw=none}
	]
	\addplot[mark size = 1.0pt, mark repeat=2, draw=blue] table [x index=0, y index=1 ,col sep=space]{Spheres_SurfaceAreaCalc.txt}; 
	\addplot[mark repeat=2, mark size = 1.0pt, draw = red] table [x index =2, y index = 3, col sep=space, ]{Spheres_SurfaceAreaCalc.txt}; 
	\addplot[ mark size = 1.0pt, only marks, draw=blue, mark = square] table [x index =4, y index = 7, col sep=space, ]{Spheres_SurfaceAreaCalc.txt};
	\addplot[only marks, mark size = 1.0pt, draw=red, mark=square] table [x index =4, y index = 10, col sep=space, ]{Spheres_SurfaceAreaCalc.txt};
	\addlegendentry{Kirchhof- 3 spheres};
	\addlegendentry{Kirchhof- 6 spheres};
	\addlegendentry{SPH- 3 spheres};
	\addlegendentry{SPH- 6 spheres};
	\end{axis}
	\end{tikzpicture}
	\caption[Dimensionless surface area evolution of agglomerate chains with three and six spheres.]{Dimensionless surface area evolution of agglomerate chains with three and six spheres. The surface area is normalized with the initial surface area $S_0$ of the primary particles. 
		The normalized surface area determined by \cite{kirchhof2009three} are given as solid lines whereas the results from are shown as hollow circles. }
	\label{Surfacearea}
  \end{center}
  \end{figure}
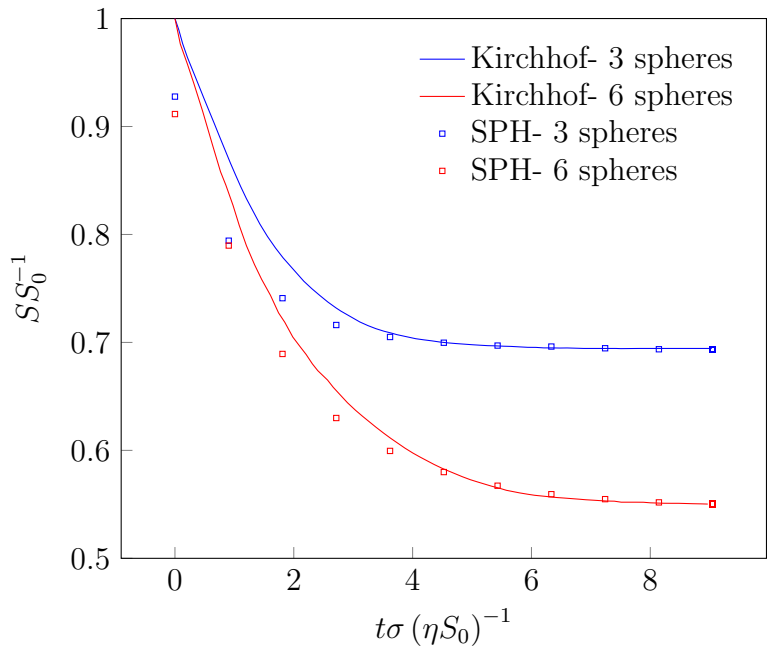

In order to motivate the application of the proposed SPH approach to 
understand structural evolution of melting solids with realistic 
transport and thermal properties, we present the 
simulation of agglomeration of melting spheres.
Sintering is sometimes approximated as a slow viscous flow \cite{kirchhof2009three} in the 
low Reynold's number regime. In such an assumption, which is indeed valid for many engineering applications, a non dimensional time can be defined as \cite{Kirchhof2009}
\begin{equation}
  \hat{t} = \frac{t\sigma}{\eta r_0},
  \label{viscoustime}
\end{equation}
where $\sigma$ is the surface tension, $\eta$ is the dynamic viscosity, $r_0$ is the characteristic length and $t$, the time. However, when the melting 
dynamics and heat transfer is fully resolved, the resulting structural change cannot be 
represented by a single non dimensional quantity. In table \ref{ThermoSMill}. we present the transport and thermal properties of the materials
we simulate. However, a high viscosity of $1 $Pa$\cdot$s is used for the 
viscous flow sintering simulation. As mentioned before, the issue of even higher viscosity is outside
the scope of the current paper and will be dealt with in a future work. 

Initially we perform simulation of viscous sintering similar to that given in \cite{kirchhof2009three}. In Fig. \ref{KirchhoffComp}, we compare the viscous
sintering results from \cite{kirchhof2009three} with SPH flow simulations. A very good
visual agreement is observed. Further, in Fig. \ref{Surfacearea}, the evolution of dimensionless surface are of the agglomerate spheres is confirmed and
the quantitative agreement is seen.

\begin{figure}[!htb]
  \begin{center}
	\begin{subfigure}[t = 0]{1.5in}
		\includegraphics[width=1.5in]{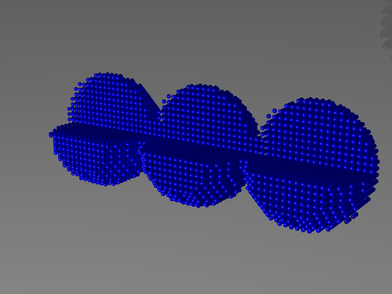}
		\caption{t=0\,s}
	\end{subfigure}
	\begin{subfigure}[]{1.5in}
		\includegraphics[width=1.5in]{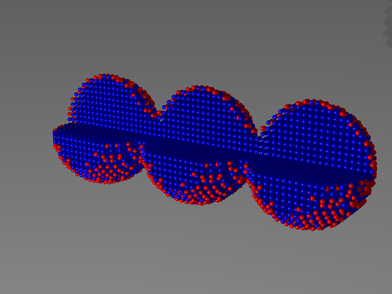}
		\caption{t=0.0025\,s}
		\label{t2}
	\end{subfigure}
	\begin{subfigure}[]{1.5in}
		\includegraphics[width=1.5in]{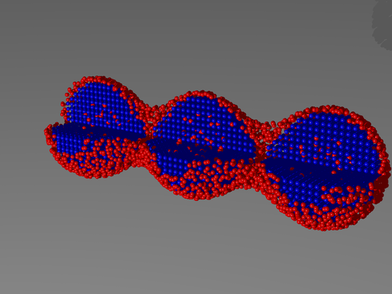}
		\caption{t=0.0089\,s}
			\label{t3}
	\end{subfigure}
	 \begin{subfigure}[]{1.5in}
	 	\includegraphics[width=1.5in]{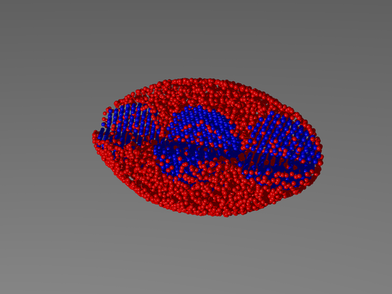}
	 	\caption{t=0.0432\,s}
	 \end{subfigure}
	 \begin{subfigure}[]{1.5in}
	 	\includegraphics[width=1.5in]{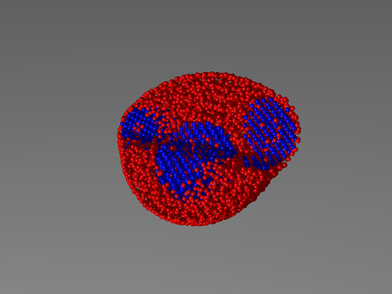}
	 	\caption{t=0.0521\,s}
	 \end{subfigure}
	 \begin{subfigure}[]{1.5in}
	 	\includegraphics[width=1.5in]{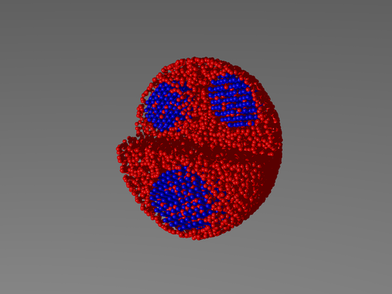}
	 	\caption{t=0.0890\,s}
	 \end{subfigure}
\begin{subfigure}[]{1.5in}
	 \includegraphics[width=1.5in]{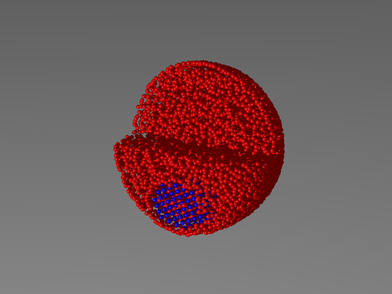}
	 \caption{t=0.1341\,s}
\end{subfigure}
\begin{subfigure}[]{1.5in}
	\includegraphics[width=1.5in]{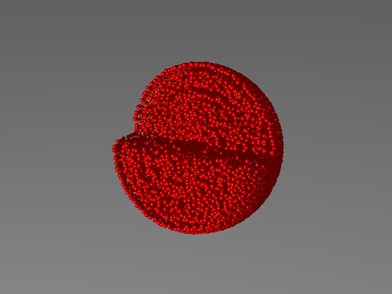}
	\caption{t=0.2084\,s}
\end{subfigure}
	\caption[{SPH} particles in solid or liquid state during the simulation of an agglomerate chain with three primary particles.]{Representation of the solid-liquid phase change of an agglomerate chain with three primary particles. Blue colored particles denote solid and red colored particles liquid state. Melting occurs until all particles are in the liquid state and form a sphere.}
	\label{agglophase}
  \end{center}
  \end{figure}

\begin{figure}[!htb]
  \begin{center}
	\begin{subfigure}[t = 0]{1.5in}
		\includegraphics[width=1.5in]{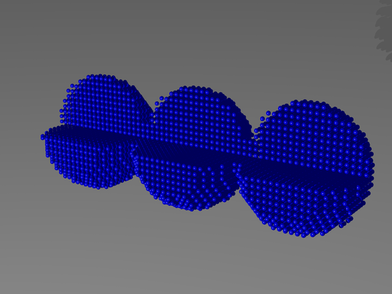}
		\caption{t=0\,s}
	\end{subfigure}
	\begin{subfigure}[]{1.5in}
		\includegraphics[width=1.5in]{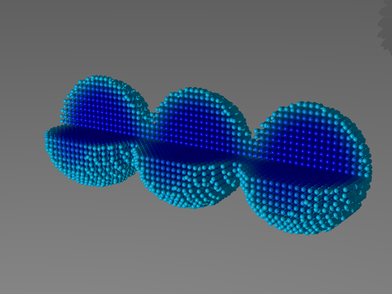}
		\caption{t=0.0025\,s}
	\end{subfigure}
	\begin{subfigure}[]{1.5in}
		\includegraphics[width=1.5in]{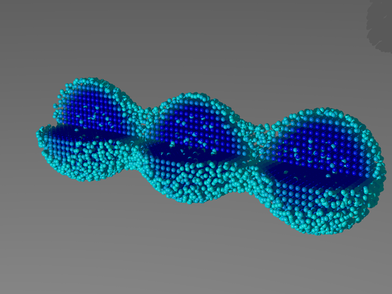}
		\caption{t=0.0089\,s}
	\end{subfigure}
	 \begin{subfigure}[]{1.5in}
	 	\includegraphics[width=1.5in]{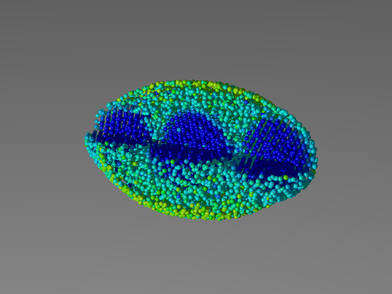}
	 	\caption{t=0.0432\,s}
	 \end{subfigure}
	 \begin{subfigure}[]{1.5in}
	 	\includegraphics[width=1.5in]{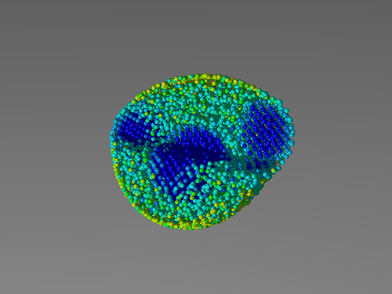}
	 	\caption{t=0.0521\,s}
	 \end{subfigure}
	 \begin{subfigure}[]{1.5in}
	 	\includegraphics[width=1.5in]{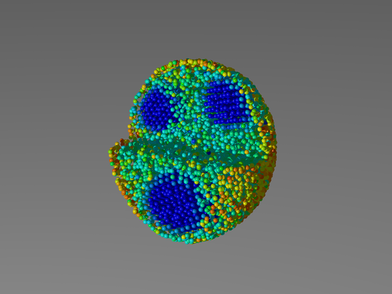}
	 	\caption{t=0.0890\,s}
	 \end{subfigure}
	 \begin{subfigure}[]{1.5in}
	 	\includegraphics[width=1.5in]{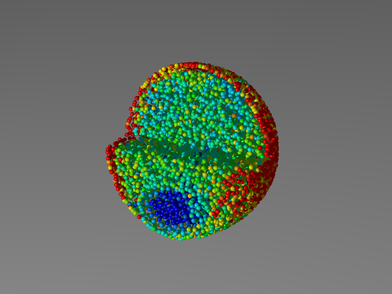}
	 	\caption{t=0.1341\,s}
	 \end{subfigure}
	\begin{subfigure}[]{1.5in}
		\includegraphics[width=1.5in]{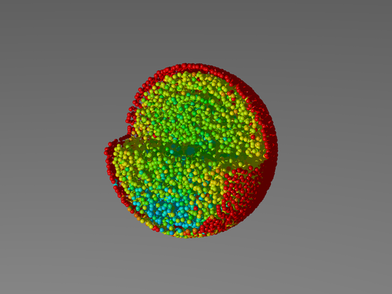}
		\caption{t=0.2084\,s}
	\end{subfigure}
	\caption[Temperature distribution at different times of an agglomerate chain with three primary particles. ]{Temperature distribution of  three primary particles forming an agglomerate chain and liquid state particles formed during the melting process. The temperature takes values between 1.0\,K (blue colored) and 3.034\,K (red colored).}
	\label{agglotemp}
  \end{center}
  \end{figure}

\begin{figure}[!htb]
  \begin{center}
	\begin{subfigure}[t = 0]{1.5in}
		\includegraphics[width=1.5in]{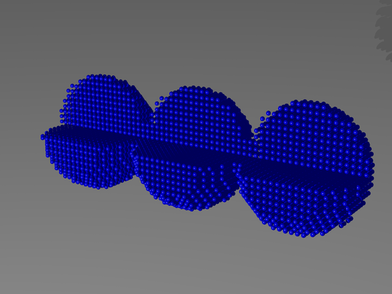}
		\caption{t=0\,s}
	\end{subfigure}
	\begin{subfigure}[]{1.5in}
		\includegraphics[width=1.5in]{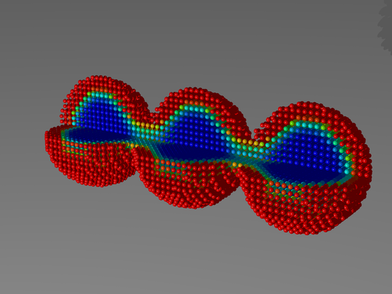}
		\caption{t=0.0025\,s}
	\end{subfigure}
	\begin{subfigure}[]{1.5in}
		\includegraphics[width=1.5in]{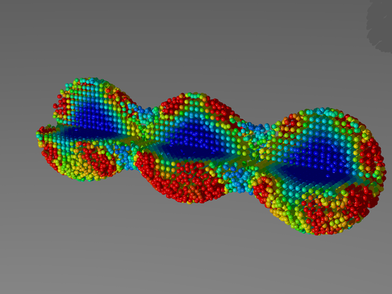}
		\caption{t=0.0089\,s}
	\end{subfigure}
	\begin{subfigure}[]{1.5in}
		\includegraphics[width=1.5in]{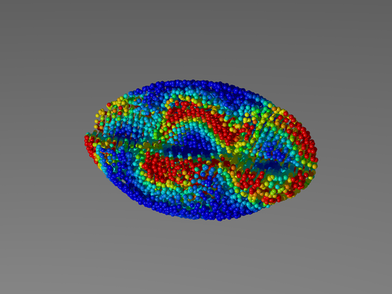}
		\caption{t=0.0432\,s}
	\end{subfigure}
	\begin{subfigure}[]{1.5in}
		\includegraphics[width=1.5in]{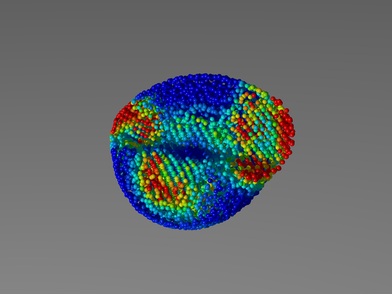}
		\caption{t=0.0521\,s}
	\end{subfigure}
	\begin{subfigure}[]{1.5in}
		\includegraphics[width=1.5in]{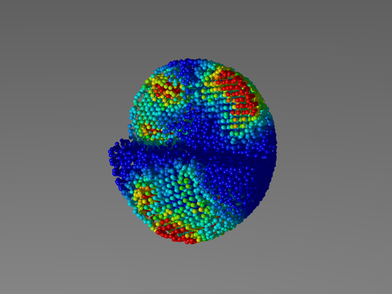}
		\caption{t=0.0890\,s}
	\end{subfigure}
	\begin{subfigure}[]{1.5in}
		\includegraphics[width=1.5in]{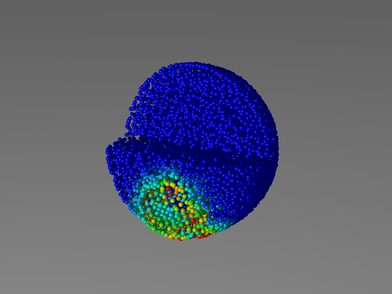}
		\caption{t=0.1341\,s}
	\end{subfigure}
	\begin{subfigure}[]{1.5in}
		\includegraphics[width=1.5in]{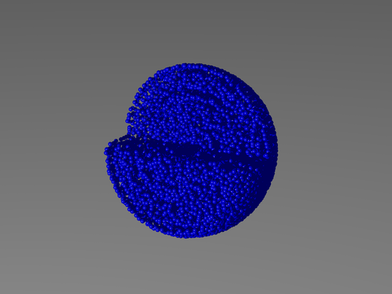}
		\caption{t=0.2084\,s}
	\end{subfigure}
	\caption[Latent heat effect at different times of an agglomerate chain with three primary particles.]{Representation of the regions in which the particles are located in the temperature range of the phase change. The effect of latent heat is normalized and has its  greatest effect in red-colored areas. Blue colored particles are outside the mushy-region for the phase change and experience no effect due to the latent heat.}
	\label{aggloL}
  \end{center}
  \end{figure}

We then present the melting dynamics of 3 spheres with finite latent heat value of $100 $ kJ$/$kg. The phase, temperature and local latent heat are 
presented in figures, \ref{agglophase}, \ref{agglotemp} and \ref{aggloL} respectively at different time instances. After $0.0025$ s, a thin layer
of liquid is formed on the spheres, which coalesce with the adjacent spheres. The resulting fluid-solid interaction results in complex 
free surface evolution, remarkably different from the assumption of
viscous sintering. Also, as seen at the time instance of $0.1341$ s, two out of the three primary spheres are molten completely, implying that the 
non linear motion of the spheres also affect heat transfer in the bulk 
of the material. 

\begin{figure}[!htb]
  \begin{center}                                                                  
  \begin{overpic}[width = 0.7\textwidth ]{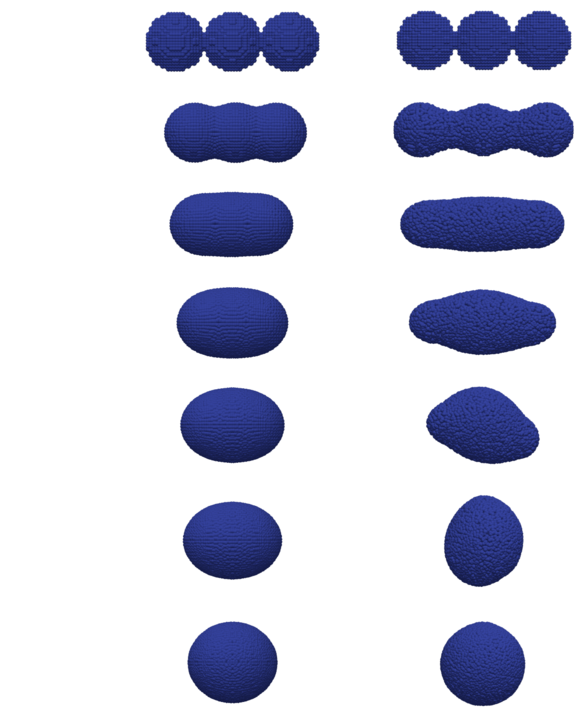}                 
    \put(0,92){$t=0\,\mathrm{s}$}                                              
    \put(0,80){$t=0.013\,\mathrm{s}$}                                          
    \put(0,66){$t=0.026\,\mathrm{s}$}                                          
    \put(0,53){$t=0.038\,\mathrm{s}$}                                          
    \put(0,38){$t=0.051\,\mathrm{s}$}                                          
    \put(0,22){$t=0.063\,\mathrm{s}$}                                          
    \put(0,08){$t=0.127\,\mathrm{s}$}                                           
  \end{overpic}                                                                 
  \caption{Comparison of the coalescence agglomerates. Left column shows the viscous flow assumption whereas the right column shows the melting process of initially solid particles.}
  \label{compareshapes}                                                                 
  \end{center}
\end{figure}

The shape evolution of dynamically melting agglomerate chain of spheres with
different latent heat capacities is then compared against the viscous flow 
assumption. Note that the viscous flow simulations have a viscosity $1000$ times larger than that of the melting simulation, while the latent heat 
of the melting spheres is $100$ kJ/kg. At these values, the shapes evolved 
at comparable rates as seen in Fig.\ref{compareshapes}. Since the fluid is not all molten for the given time instances, 
the fluid-solid interaction results in complex shapes, often resulting in 
low aspect ratio shapes compared to the result from the linear assumptions.

\begin{table}
		\centering
	\caption{Thermodynamic properties of the solid and liquid particles as well as the ambient pseudoparticles for the melting of an agglomerate chain with three primary spheres. }
	\label{ThermoSMill}
	\begin{tabular}{c c c }
		\toprule
		Quantity& Value& Unit \\
		\midrule
			$\sigma$ & $7.12\cdot 10^{-2}$ & $\mathrm{N}\,\mathrm{m}^{-1}$\\
				$\Theta$ & 30.00& $^{\circ}$\\
				$\eta$	& $1.00\cdot 10^{-3}$  & $\mathrm{Pa}\,\mathrm{s}$\\
		$\rho $ & $1.00\cdot 10^{3}$& $\mathrm{kg}\,\mathrm{m}^{-3}$\\
	$T_0$ 		& $1.00$		&		$\mathrm{K}$				\\
		$T_{\mathrm{m}}$ & 	1.15 &	$\mathrm{K}$					\\
	$k_{\mathrm{s}}$ & 	$	2.14$		&	$\mathrm{W}\,\mathrm{m}^{-1}\,\mathrm{K}^{-1}$			\\
	$k_{\mathrm{l}}$ & 		$	0.56$	&	$\mathrm{W}\,\mathrm{m}^{-1}\,\mathrm{K}^{-1}$		\\
	$c_{p,\,\mathrm{s}}$ & 	$2.11\cdot 10^{3}$	&	$\mathrm{J}\,\mathrm{kg}^{-1}\,\mathrm{K}^{-1}$			\\
	$c_{p,\,\mathrm{l}}$ & 	 	$4.22\cdot 10^{3}$ &	$\mathrm{J}\,\mathrm{kg}^{-1}\,\mathrm{K}^{-1}$			\\ 
	$L$ & $1.00	$		&$\mathrm{J}\,\mathrm{kg}^{-1}$	\\
		$T_{\mathrm{amb}}$ &	4.00	&	$\mathrm{K}$			\\
	$k_{\mathrm{amb}}$ & 		0.56&	$\mathrm{W}\,\mathrm{m}^{-1}\,\mathrm{K}^{-1}$			\\	 
	$c_{p,\,\mathrm{amb}}$ &  	$4.22\cdot 10^{3}$	&	$\mathrm{J}\,\mathrm{kg}^{-1}\,\mathrm{K}^{-1}$		\\ 
	\bottomrule
	\end{tabular} 
\end{table}


\section{Conclusions}
A multiphysics, coupled heat transfer, phase change and capillary flow
solver is presented in the context of Incompressible Smoothed Particle Hydrodynamics. 
A constant volume melting model is implement. The model is based on a spefic heat
that increases additively by the latent heat of the material, near the melting 
interface and at temperatures close to the melting point. Heat transfer 
accross the free surface through a semi analytic Dirichlet BC across the free surface
allows the study of complex shaped bodies. The method is 
carefully validated against theoretical models on melting and conduction heat 
transfer. 

The method is first applied to study the influence of particle shape on it 
surface evolution during heating and subsequent melting of the particle. 
We simulate melting process across a range of latent heat values and show that
at high latent heat values the melting process evolves substantially differently 
from assumptions of a spherical shape. A non spherical shape causes capillary 
flow of the melt, increasing heat transfer due to convection within the melt 
accelerating the melting process. 

The method is then applied to the study of agglomeration of a chain of 
particles that undergo melting. The results are compared against simulations 
that assume viscous sintering. We show that when unmolten solid particles are 
present (owing to large latent heat of melting of the material), the evolution
of shape of the agglomerate follows a highly non linear path due to the fluid-solid
interaction within the body of the melt. Thus a case is made for the method's 
application to microstructure studies in additive manufacturing where time scales 
of the flow and phase change are comparable. 

In this work, we have used representative transport and thermal properties. However
this doesn't span the entire range of properties encountered in real applications. 
For example high viscosities encountered in additive manufacturing using polymers
would require implicit computation of viscosities. Simulations with high latent 
heat require adaptive time stepping to reduce computational costs. These improvements
in computational efficiency will enable simulations of large number of particles
needed for relating microstructure to bulk properties of manufactures parts. 

\clearpage
\newpage
\bibliography
\clearpage
\newpage
\end{document}